\documentclass{aa}    

\usepackage{graphicx}
\usepackage[mathcal]{eucal}
\usepackage{amssymb}
\usepackage{amsmath}
\usepackage{natbib}
\usepackage[T1]{fontenc}

\def\gds {$\gamma$ Dor stars}
\def\dss {$\delta$ Scuti stars}
\def\ds {$\delta$ Scuti star}
\def\vsini {{v\!\sin\!i}} 
\def\kms {{\mathrm{km}\,\mathrm{s}^{-1}}}
\def\dst {\displaystyle}
\def\teff {{T_{\mathrm{eff}}}}
\def\msol {{\mathrm{M}_\odot}}
\def\sun {{_{\odot}}}
\def\rsol {{R\sun}}

\def\muHz {{\mu\!\mbox{Hz}}}

\def\bs {\boldsymbol}
\newcommand{\inv} {\frac {1}}
\def\msol {{\mathrm{M}_\odot}}

\def\rhoo {{\rho_0}}
\def\rhod {{\rho_2}}

\newcommand{\rotbar}{{\bar \Omega}}

\newcommand{\derivplnln} [2] {\frac {\partial\ln #1 } {\partial\ln #2} }

\newcommand{\deriv} [2] {\frac {d #1 } {d #2} }

\newcommand{\kro} [2] {\delta_{#1#2}}

\def\utier {\frac{1}{3}}

\def\udemi {\frac{1}{2}}

\def\cost {{\cos\theta}}
\def\cosdt {{\cos^2\!\theta}}

\def\i {{\emph{i}}}
\def\j {{\emph{j}}}

\def\dst {{\displaystyle}}

\def\la {{\ell_a}}
\def\lb {{\ell_b}}

\def\lbpd {{\ell_{b+2}}}
\def\lbmd {{\ell_{b-2}}}

\def\etao {{\eta_0}}
\def\etad {{\eta_2}}

\def\aet {{A^*}}

\def\ud {{u_2}}
\def\bd {{b_2}}
\def\bt {{b_3}}
\def\du {{d_1}}
\def\dd {{d_2}}

\def\vg {{V_{\mathrm{g}}}}

\def\gamu {{\Gamma_1}}
\def\po {{p_0}}

\def\Eu {{E_1}}
\def\Ed {{E_2}}
\def\Et {{E_3}}
\def\Eq {{E_4}}

\def\sa {{s_a}}
\def\sb {{s_b}}

\def\qa {{q_a}}
\def\qb {{q_b}}

\def\ysab {{\left(\fa\sb+\fb\sa\right)}}

\def\intr {{\int_0^{\mathrm{R}}}}
\def\dr {{\,\mbox{d}r}}

\def\intang {{\int_{\underline{\Omega}}}}

\def\dang {{\,\mbox{d}\underline{\Omega}}}

\def\qdab {{\mathcal Q}_{2ab}}
\def\zab {{\mathcal Z}_{ab}}

\def\fa {{y_a}}          
\def\yoda {{y_{02,a}}}
\def\va {{v_a}}
\def\wa {{W_a}}
\def\fb {{y_b}}         
\def\yodb {{y_{02,b}}}
\def\vb {{v_b}}
\def\wb {{W_b}}

\def\you {{y_{01}}}
\def\zo {{z_{0}}}
\def\yod{{y_{02}}}
\def\yot{{y_{03}}}
\def\yoc{{y_{04}}}

\def\yu  {{y_1}}
\def\zu  {{z_1}}
\def\yd  {{y_2}}
\def\yt  {{y_3}}
\def\yc  {{y_4}}

\def\yub {{y_{\mathrm{1b}}}}

\def\zub {{z_{\mathrm{1b}}}}
\def\youa {{y_{\mathrm{01,a}}}}
\def\youb {{y_{\mathrm{01,b}}}}
\def\zoa {{z_{\,0,a}}}
\def\zob {{z_{\,0,b}}}
\def\geff {{g_{\mathrm{eff}}}}

\def\sigocar {{\sigma_0^2}}

\def\wua {{\omega_{1,a}}}

\def\wub {{\omega_{1,b}}}

\def\w   {{\omega}}
\def\wc  {{\hat \omega}}
\def\wo  {{\omega_0}}
\def\wod {{\omega^2_0}}

\def\wu  {{\omega_1}}

\def\woj {{\omega_{0,\j}}}  

\def\wok {{\omega_{0,k}}}  
\def\wuj {{\omega_{1,\j}}}  
\def\wuk {{\omega_{1,k}}}  
\def\wut {{\tilde{\omega}_1}}
\def\wdt {{\tilde{\omega}_2}}

\def\wuab  {{\w_{\mathrm{1,ab}}}}             

\def\wujk  {{\w_{\mathrm{1,\j k}}}}     
\def\wukj  {{\w_{\mathrm{1,k \j}}}}     
\def\wdD  {{\w_{\mathrm{2}}^{\mathrm{D}}}} 
\def\wdjD  {{\w_{\mathrm{2,\j}}^{\mathrm{D}}}}   
\def\wdabD {{\w_{\mathrm{2,ab}}^{\mathrm{D}}}}
\def\wdabI {{\w_{\mathrm{2,ab}}^{\mathrm{I}}}} 

\def\wdjkD {{\w_{\mathrm{2,\j k}}^{\mathrm{D}}}}  
\def\wdT   {{\w_{\mathrm{2}}^{\mathrm{T}}}}

\def\wdjT   {{\w_{\mathrm{2,\j}}^{\mathrm{T}}}}  
\def\wdjP   {{\w_{\mathrm{2,\j}}^{\mathrm{P}}}}  
\def\wdjI   {{\w_{\mathrm{2,\j}}^{\mathrm{I}}}}  
\def\wdP   {{\w_{\mathrm{2}}^{\mathrm{P}}}}
\def\wdabT {{\w_{\mathrm{2,ab}}^{\mathrm{T}}}}

\def\wdjkT {{\w_{\mathrm{2,\j k}}^{\mathrm{T}}}}  
\def\wdjkP {{\w_{\mathrm{2,\j k}}^{\mathrm{P}}}}  
\def\wdjkI  {{\w_{\mathrm{2,\j k}}^{\mathrm{I}}}}
\def\wdI  {{\w_{\mathrm{2}}^{\mathrm{I}}}} 

\def\wuba  {{\w_{\mathrm{1,ba}}}}             

\def\wdabPT {{\w_{\mathrm{2,ab}}^{\mathrm{PT}}}}

\def\wdabP {{\w_{\mathrm{2,ab}}^{\mathrm{P}}}}

\def\mua {{\mu_{a}}}
\def\mub {{\mu_{b}}}
\def\muab {{\mu_{ab}}}
\def\muba {{\mu_{ba}}}
\def\mujk {{\mu_{\j k}}}
\def\mukj {{\mu_{k \j}}}
\def\muj {{\mu_{\j}}}
\def\muk {{\mu_{k}}}

\def\xig   {{\bs {\xi}}}
\def\xioa  {{\bs \xi_{\mathrm{0,a}}}}

\def\xioabra {{\langle\xioa|}}      

\def\xiob  {{\bs \xi_{\mathrm{0,b}}}}
\def\xiobket {{|\,\xiob\rangle}}      

\def\xiubpket {{|\,\xiubp\rangle}}      
\def\xiubtket {{|\,\xiubt\rangle}}      

\def\xio {{\bs \xi}_{0}}

\def\xiu {{\bs \xi}_{1}}

\def\xiukP {{\bs \xi_{1,k}^{P}}}  
\def\xiukT {{\bs \xi_{1,k}^{T}}}  

\def\xiubp {{\bs \xi}^{\mathrm{P}}_{\mathrm{1,b}}}

\def\xiubt {{\bs \xi}^{\mathrm{T}}_{\mathrm{1,b}}}

\def\ylm {{Y_\ell^m}}

\def\ylmp {{Y_{\ell+1}^m}}

\def\ylmm {{Y_{\ell-1}^m}}

\def\yac {{Y_a^*}}

\def\yb {{Y_b}}

\def\Fl  {{\mathcal F}_{\ell}}
\def\Fb  {{{\mathcal F}_b}}

\def\elo {{\bs {\cal L}_0}}
\def\eld {{\bs {\cal L}_2}}
\def\elos {{\bs L}_0}
\def\elds {{\bs L}_2}
\def\ka  {{\bs {\cal K}}}
\def\kas {{\bs K}}

\def\gradh {{\bs\nabla_{\mathrm{H}}}}

\def\mjk {{{\cal M}_{\j k}}}

\def\ez  {{\bs e_z}}
\def\er  {{\bs e_r}}

\def\gamb {{\gamma_b}}
\def\gambp {{\gamma_{b+1}}}

\def\gama {{\gamma_a}}

\def\gamap {{\gamma_{a+1}}}

\def\delb {{\Lambda_{b}}}
\def\delbp {{\Lambda_{b+1}}}

\def\dela {{\Lambda_{a}}}

\def\delabp {{\Lambda_{ab}^{+}}}

\def\delabm {{\Lambda_{ab}^{-}}}
\def\delbam {{\Lambda_{ba}^{-}}}

\def\lambb {{\lambda_b}}

\newcommand{\eqn} [1] {
\begin{equation}
#1
\end{equation}}

\newcommand{\eqna} [1] {
\begin{eqnarray}
#1
\end{eqnarray}}

\def\geff {{g_{\mathrm{eff}}}}

\def\Ux {{U_{\chi}}}
\def\dg {{DG92}}
\def\soufi {{SGD98}}
\def\gp {{$gp\,$}}

\begin{document}


   \title{Effects of moderately fast shellular rotation on adiabatic
                 oscillations}
   \titlerunning{Effects of shellular rotation on adiabatic
                 oscillations}
   \authorrunning{Suárez, Goupil \& Morel}
   
   \author{J.C. Su\'arez\inst{1,2}
             \thanks{Associate researcher at institute (2), with finanantial 
	      support from Spanish <<Consejería de Innovación, Ciencia 
	      y Empresa>> from the <<Junta de Andalucía>> local government.}
	      \and M.J. Goupil\inst{2} 
	      \and P. Morel\inst{3} } 

   \offprints{J.C. Su\'arez\,\email{jcsuarez@iaa.es}}

  \institute{Instituto de Astrofísica de Andalucía (CSIC), CP3004, Granada,Spain               
  \and LESIA, UMR 8109, Observatoire de Paris-Meudon, France 
  \and Observatoire de la Côte d'Azur, Nice, France}

   \date{Received un jour; Accepted un autre jour}

   \abstract{We investigate adiabatic oscillations
             for $\delta$ Scuti star models, taking into account a
	     moderate rotation velocity ($\sim100\,\kms$).
	     The resulting oscillation frequencies include corrections for  rotation
             up to second order in the rotation rate
	     including those of near degeneracy. Effects of either a uniform rotation or
	      a rotation profile assuming local angular momentum conservation of the form
	     $\Omega=\Omega(r)$  on oscillation frequencies are compared.
	      As expected, important differences (around $3\,\muHz$) are
	     obtained in the $g$ and mixed mode regions. For  higher frequency $p$ modes,
	     differences range between $1\,\muHz$ and $3\,\muHz$. Such differences are likely to be
	     detectable with future space missions such as COROT, where
	     precisions in frequency around $0.5\,\muHz$ will be reachable.
	     
             }

   \maketitle

\section{Introduction\label{sec:Intro}}

Intermediate mass stars  are characterized by a convective
core and a radiative envelope. As representative of such stars, \dss\ are
located in the lower part of the Cepheid instability strip, with spectral
types from A2 to F0. Such pulsating stars seem particularly suitable
for determining the extent of the convective core and internal 
rotation rates, and thereby  to understand better hydrodynamical processes
taking place  in stellar interiors. Particularly, balance between rotationally 
induced turbulence and meridional circulation generates mixing of chemicals 
and redistribution of angular momentum \citep{Zahn92}, which affects the 
rotation profile and the evolution of the star. It is expected that intermediate 
mass stars do not rotate uniformly as a solid body. \citet{Zahn92} proposed 
that, as a result of strong anisotropic turbulence, the stellar rotation profile 
is \emph{shellular}, i.e. the star is divided in differentially rotating 
concentric shells.

\dss\ can be found  in two principal evolutionary stages: on the main 
sequence, where they burn hydrogen in their convective cores, and before the 
sub-giant phase, burning hydrogen in shells. They can be sufficiently evolved 
that they  present a strong gradient of chemical composition, produced by their 
shrinking convective core. These types of structures generate particular modes, 
showing a \emph{dual} behaviour. These modes are known as \emph{mixed} modes, 
which behave as pressure modes ($p$ modes) at surface, and as gravity modes 
($g$ modes) toward the center \citep{JCD98}. {\bf Indeed, they
can penetrate sufficiently deep toward the center of the star 
and at the same time show amplitudes at the surface large enough to be detected. 
The existence of such modes can thus be 
very important when investigating deep interiors of stars}.

In the last decades,  great efforts have been employed to develop observational 
seismology of \dss, for instance the World-wide observational campaigns 
\citep{Breger00,Handler00} or \emph{STEPHI} network \citep{Michel00stephi}. However, 
several observational aspects of the pulsating behaviour within the instability strip 
are not fully understood \citep[see][]{Templeton97}. Due to the complexity of the 
oscillation spectra of \dss, the problem of the identification of modes remains unsolved. 
As \dss\ are commonly fast rotators ($100\!<\!\vsini\!<\!200\,\kms$), additional 
uncertainties to their already complex oscillation spectra arise from the effect 
of rapid rotation. Rotation alters the internal structure of a star through 
a modified  hydrostatic balance and, likely more important, through mixing caused by 
circulation and/or instabilities induced by rotation 
\citep{Zahn92, MaederMeynet00, Heger00}. Furthermore the simple characteristic pattern 
of symmetric multiplets split by rotation is broken. In the framework of a perturbation 
analysis, second-order effects induce strong asymmetries in the splitting of multiplets 
\citep[][hereafter \dg]{Saio81, DG92} and frequency shifts which cannot be neglected even 
for radial modes \citep{Soufi95}.

In \dg , the authors propose a second-order formalism to study the eigenfunctions for 
the Sun and \dss, with both radial and latitudinal rotation, which takes into account
the effects of near degeneracy  up to first order in the rotation rate. For medium-high 
rotators, like \dss, the effects of near degeneracy up to second and third order cannot 
be neglected \citep[see][]{DG92,Goupil00}. This last point is included in a complete 
third order formalism given by \citet{Soufi98}, hereafter \soufi. Near degeneracy affects 
the asymmetry of multiplets, making  even  more difficult the interpretation of observed 
oscillation spectra. Based on this approach, assuming a uniform rotation for the sake of 
simplicity, theoretical and quantitative studies on oscillations of rotating $\delta$ Scuti 
stars \citep{Goupil00,GoupilTalon02,Pagoda02, Alosha03}; on solar-like stars 
\citep{DziembowskiGoupil98, Goupil04}, and $\beta$~Cephei stars \citep{Pagoda03bceph} have 
followed. \soufi's work has been recently revisited and applied to the study 
of oscillations of $\beta$~Cephei stars \citep[][ in press]{Karami05}.

The present work focuses on the magnitude of the effect of shellular rotation on adiabatic 
oscillations of a $1.8\,\msol$ star. In order to avoid possible \emph{interferences} 
between third order effects of rotation on oscillations and those coming from shellular 
rotation, only second-order terms are considered. To do so, we have built a numerical 
code \citep{SuaThesis} taking into account a complete formalism up to second order in 
presence of a shellular rotation $\Omega=\Omega(r)$.

Theoretical oscillation spectra computed for two different models of $\delta$ Scuti star 
are compared: one model has been evolved assuming  uniform rotation (global conservation 
of the angular momentum), and another model has been evolved assuming a shellular rotation 
(local conservation of the angular momentum). The impact of a shellular rotation on 
oscillation frequencies is discussed in the frame of the future  space experiment COROT 
(launch in 2006).

The paper is organized as follows: Sect.~\ref{sec:OscFreq} briefly reminds the basis of 
the second-order perturbation formalism of non-degenerate oscillation frequencies. 
Section~\ref{sec:neardeg} describes near degeneracy theory as implemented in our 
oscillation code. In Sect.~\ref{sec:evolmodels} the adopted methodology is detailed. 
Section~ \ref{sec:effofDR} discusses the results and finally conclusions are given in 
Sect.~\ref{sec:conclu}.

\section{Non-degenerate oscillation frequencies corrected for rotation effects
         \label{sec:OscFreq}}

Following \dg, when Coriolis and centrifugal forces are considered,
the eigenfrequency $\omega$ and associated eigenfunctions 
$\xig$ must satisfy the following oscillation equation:
  \eqn{\elo\xig-\rhoo\wc^2-2 \rhoo \wc \Omega  \kas \xig+(\eld-\rhod\wc^2)\,\xig=0,
  \label{eq:osc}}
where $\wc=\w+m \Omega$ and $\kas=\i\ez\times$ (\soufi, Eq.~22). The linear
operators $\elo$ and $\eld$ respectively correspond to 
$\elos$ and $\elds$ defined in \soufi\ (Eqs.~23 and 24).
The mean density of the pseudo rotating model is represented by $\rho_0$. 
Respectively, its perturbation by the non-spherically symmetric component of the 
centrifugal force is given by the following expansion in Legendre polynomials
\eqn{\rho_2 = p_{22}(r)\,P_{2}(\cos\theta),\label{eq:defrho2}}
where $p_{22}(r)$ is defined in \soufi\ (Eq.~15). With appropriate boundary 
conditions, Eq.~\ref{eq:osc} forms an eigenvalue problem 
(appendix \ref{ap:oscil-pseudo}). The eigenfunction $\xig$ is written as 
\eqn{\xig = \xio +\xiu\,,} 
where $\xi_0$ and $\xi_1$ respectively correspond to the non perturbed and first-order 
perturbed eigenfunctions, defined as:
\eqna{\xio&=& r\Big[\you\ylm\er+\zo\gradh\ylm\Big]\label{eq:defxio}\\
      \xiu&=& \frac{2m\rotbar}{\wo}\,r\,\Big[\yu\ylm\er
               +\zu\gradh\ylm\Big.\nonumber\\
               &+&\,\Big. \tau_{\ell+1} \, \er\times\gradh\ylmp+
      \hat \tau_{\ell-1} \,  \er\times\gradh\ylmm\Big]\,,\label{eq:defxiu}}
where the notations are the same as in \soufi.
In these expressions, $\you$ and $\zo$ respectively represent the normalized 
radial and horizontal components of the eigenfunction for the fluid displacement
\citep{Unno89}. Similarly, the first-order corrections are represented by $y_1$ 
and $\zu$ respectively (see Appendix~\ref{sap:1stordEigenf}).

Hence, the eigenfrequency associated with each mode is labeled with the subscripts 
$n, \ell, m$ representing, respectively, the radial order, degree and, 
azimuthal order of the corresponding spherical harmonics. The temporal dependence 
is chosen of the form $exp(i\omega t)$, so that prograde modes correspond to $m < 0$. 
Following \dg\, oscillation frequencies are obtained by means of a perturbative 
method taking into account up to second order effects of rotation. Equilibrium and 
oscillating quantities are expanded with respect to $\epsilon=\Omega\,\sigma/\omega$ 
and $\mu=\Omega/\sigma$, where $\sigma$ is the dimensionless mode frequency defined 
as $\sigma = \omega/(GM/R^3)^{1/2}$; $R$ and $M$ represent the stellar radius and 
the mass of the model respectively.

Ignoring in this section the resonant interaction due to near degeneracy, the frequency 
of a given rotationally split multiplet of degree $\ell$ and radial order $n$ can be 
written as:
\eqn{\omega_m = \omega_0 + \omega_{1,m} + \omega_{2,m},\label{eq:orders012}}
where the subscripts $n, \ell$ have been omitted. The $\omega_m$ frequencies
are obtained by means of a perturbative method. The zeroth-order contribution, $\omega_0$, 
represents the mode frequency with all effects of rotation ignored except that of the 
horizontally averaged centrifugal force in the equilibrium model (later on, we will 
abusively refer to it as the `non perturbed' frequency). The other two terms, 
$\omega_{1,m}$ and $\omega_{2,m}$ represent the first- and second-order corrections, 
respectively.

A shellular rotation is considered with a rotation profile defined as 
\eqn{\Omega(r)= \rotbar \, \dst[1+\eta_0(r)]\,, \label{eq:defeta0}}
where $\rotbar$ represents the rotation frequency at the stellar surface.
In this context, the first- and second-order frequency corrections are
written as:
\eqna{\omega_{1,m} & = & m\,\rotbar \, (C_{L}-1-J_0)\label{eq:w1m} \\
      \omega_{2,m} & = & \frac{\rotbar^2}{\omega_0}\label{eq:w2m} \,
      \left(D_0+m^2D_1\right)\,.\label{eq:w2tot_saio}}
The Ledoux constant, $C_L$, which determines the usual equidistant splitting
valid in the limit of slow rotation, is given by the well-known integral
expression:
\eqn{C_L =\inv{I_0}\int_0^R [2\you\zo+\zo^2]\,\rho_0\,r^4\,dr\,,
     \label{eq:defCl}}
where $I_0$ is the term of inertia given by:
\eqn{I_0 = \int_0^R [\you^2+\Lambda\zo^2]\,\rho_0\,r^4\,dr.\label{eq:defI0}}
with $\Lambda = \ell(\ell+1)$. In Eq.~\ref{eq:w1m}, $J_0$ represents an additional 
contribution in the case of shellular rotation and is given by:
\eqn{J_0 =\inv{I_0}\int_0^R \etao(r)[\you^2+\Lambda\zo^2-2\you\zo-\zo^2]\,
           \rho_0\,r^4\,dr\,.\label{eq:defJ0}}
The second-order coefficients $D_0$ and $D_1$ in Eq.~\ref{eq:w2tot_saio}
take into account the non-spherically symmetrical distortion due to the centrifugal 
force. The symmetry of split multiplets is broken by the $m^2$ dependency
(see Eq.~\ref{eq:w2tot_saio}). For later use, we also rewrite these coefficients in 
Saio's notation:
\eqn{D_0 \equiv X_1 + X_2 ~~~~~~~~~     D_1 \equiv Y_1 + Y_2 \label{eq:d1d2saionot}}
Comparing Eq.~\ref{eq:w2tot_saio} with the formulation
given in \dg, the following relations are obtained:
\eqn{\wo\left(\frac{\rotbar}{\wo}\right)^{2}\,(X_1+m^2Y_1)=
     \wdT+\wdI+\frac{\omega_1^2}{2\wo}+\wdP\,,}
for $X_1$ and $Y_1$ and,
\eqn{\wo\left(\frac{\rotbar}{\wo}\right)^{2}\,(X_2+m^2Y_2)
      = \wdD\label{eq:w2D}}
for $X_2$ and $Y_2$. The analytic expressions for $X_i$ and $Y_i$ are given in 
appendix \ref{ap:saioform}. The terms $\wdT$, $\wdI$ and $\wdP$, defined by 
Eqs.~16--19 in \dg, represent the different contributions to second-order frequency 
corrections in absence of degeneracy effects and, as in \soufi, include the effect 
of the symmetrical component of the centrifugal force on the equilibrium model:
$\wdP$ and $\wdT$ represent, respectively, the poloidal and toroidal components of the 
frequency, obtained from the first-order correction to the eigenfunction; $\wdD$ 
corrects for the effects of the centrifugal force on the stellar structure.

\section{Near-degenerate oscillation frequencies\label{sec:neardeg}}

The perturbation method presented in the previous section assumes that the 
non-perturbed eigenmode can be represented with one single spherical harmonic. 
This is no longer valid in the case of near-degenerate frequencies, i.e. when 
two or more frequencies are close to each other ($\omega_{nlm}\sim\omega_{nlm'}$). 
Consequently, the perturbation method must be modified in order to include 
corrections for near degeneracy as done in the next sections.

\subsection{The oscillation equation in presence of degeneracy 
           \label{ssec:EOenDegen}}

The formalism is similar for 2 or 3 near-degenerate modes. For the sake of simplicity, 
the formalism is illustrated below for two modes $a$ and $b$. The $a$ and $b$ 
subscripts represent $(n,\ell,m)_a$ and $(n,\ell,m)_b$ respectively. For near 
degenerate mode $a$ and mode $b$, it is convenient to define:
\eqna{\bar\omega_0&=& \frac{\omega_{0,a}+\omega_{0,b}}{2}\label{eq:omegabar}\\
     \delta \omega_0  &=& \omega_{0,a}-\omega_{0,b}\,.
\label{eq:defxi0}}
The eigenfrequency and the eigenfunction of a near-degenerate mode are then 
assumed of the form:
\eqna{\w&=&\bar \omega_{0}+\wut+\wdt \label{eq:defw}\\
      \xig &=&\sum_{j=a,b}\alpha_j (\xi_{0,j}+\xi_{1,j})\,.
\label{eq:defxi}}
First and second-order corrections to the eigenfrequency in presence of near 
degeneracy are represented by $\wut$ and $\wdt$ respectively; 
$\xi_{0,j}$ and $\xi_{1,j}$ are the non perturbed and first-order perturbed 
eigenfunctions respectively defined in Eq~\ref{eq:defxio} 
\& Eq.~\ref{eq:defxiu}.

Let us now replace $\w$ and $\xig$ by Eq.~\ref{eq:defw} and Eq.~\ref{eq:defxi} 
in the oscillation equation (Eq.~\ref{eq:osc}). Perturbation is then performed 
keeping in mind that $ \delta \omega_0 $ is small, of first or second order in 
the rotation rate i.e. $\delta \omega_0 \sim O(\Omega)$ or 
$\delta \omega_0 \sim O(\Omega^2)$. Projecting onto the non-perturbed eigenfunctions 
$\xioa$ and $\xiob$, the following system is obtained:
\eqn{\sum_{j=a,b} \left[{\cal M}_{jk}^{(1)}+{\cal M}_{jk}^{(2)}+
\frac{\delta \omega_0}{2}{\cal I}_{-1}+ \frac{\delta \omega_0^2}{8 \omega_0}{\cal I}_{1}\right] \, 
\alpha_{j} =0 \label{eq:defsys-2}}
for $k=a,b$. In this equation we have defined the unit matrix ${\cal I}_1$ and 
also the matrix ${\cal I}_{-1}$ as:
\eqna{{\cal I}_{-1} &=&\begin{pmatrix}
 1 & 0 \\
 0 & -1
\end{pmatrix}\,.\label{eq:mjk3}}
The first- and second-order matrix ${\cal M}_{jk}$ terms are given by:
\eqna{{\cal M}_{jk}^{(1)}&=&\begin{pmatrix}
 -\wut+\wuj & \wujk \\
  \wukj & -\wut+\wuk
\end{pmatrix}\label{eq:mjk1}\\
{\cal M}_{jk}^{(2)}&=&\begin{pmatrix}
 -\wdt+\muj & \mujk \\
 \mukj &  -\wdt+\muk 
\end{pmatrix},\label{eq:mjk2}}
where we have defined:
\eqna{\muj&=&-\frac{\wut}{\woj}\left(\frac{\wut}{2}-\wuj\right)+
      \wdjD+\wdjP+\wdjT+\wdjI \label{eq:Defmuj}\\
      \mujk\!\!&=&\wut\frac{\wujk}{\wok}+\wdjkD+\wdjkP+\wdjkT+\wdjkI\,.
      \label{eq:Defmujk}}
Terms with one single subscript are not affected by near degeneracy and are 
defined in Sect.~\ref{sec:OscFreq}. Coupling terms for degeneracy are 
included in  $ \wujk$ and $ \mujk$; they are defined with double-subscript 
terms and their definitions and detailed expressions are given in 
Appendix~\ref{ap:terms2ord}.
\begin{figure}
   \hspace{-1cm}
   \includegraphics[width=10cm]{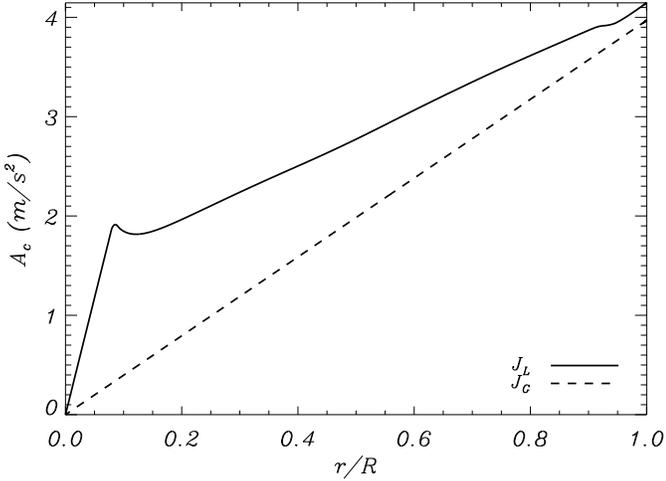}
   \caption{Radial component of the centrifugal acceleration for a 
           differentially rotating stellar model with $J_L$ 
	   (continuous line), and a uniform rotating one with $J_G$ 
	   (dashed line), considered in the present study. These 
	   quantities are displayed for a $1.8\,\msol$ model, aged 
	   of 1050 Myr with a photometric radius of $R=2.27\rsol$.}
   \label{fig:ComparAcelCentrif}
\end{figure}
\subsection{Selection rules \label{ssec:SR}}

Corrections for near degeneracy exist only for modes with degrees $\ell$ and 
azimuthal orders $m$ fulfilling specific selection rules: 
\begin{itemize}

  \item[$\bullet$] The analysis of $\wuab$ (Eq.~\ref{ap.eq:solwuab}) shows 
                   that frequency corrections for near degeneracy already arise 
		   at first order whenever the modes have same degree $\ell$ and 
		   same azimuthal order $m$, i.e. $\ell_a=\ell_b$ and $m_a=m_b$. 
		   The modes then differ only by their radial orders $n_a\not= n_b$.

  \item[$\bullet$] Near degeneracy affects second-order frequency corrections of 
                   near-degenerate frequencies (see Appendix~\ref{ap:terms2ord}) 
		   whenever $\la=\lb $ or $\la=\lb\pm2$ and $m_a=m_b$. These 
		   selection rules are also obtained in the case of the 
		   \emph{compound} third order treatment given by \soufi. 
                   Eq.~\ref{eq:defsys-2} is then solved for $j=a$ and $k=b$ for 
		   each case allowed by the selection rules.

\end{itemize}

For two near-degenerate modes, $a$ and $b$, the proximity in frequency is expected 
to be less or equal to the rotation frequency of the stellar model 
($|\omega_{a}-\omega_b|\lesssim \Omega $ or $\Omega^2/\omega_a$) depending on the 
degrees and azimuthal orders of the modes. This estimate may vary with the 
nature of the modes ($g$ modes, $p$ modes or mixed modes). For stellar models of 
our interest here, numerical applications reveal that many modes have 
near-degenerate frequencies.

\subsection{First-order near degeneracy\label{ssec:PremOrdDeg}}

In this case $\delta \omega_0$ is $O(\Omega)$. Only modes $a$ and $b$ with 
$\ell_a=\ell_b$ modes are affected. Such a situation generally concerns only a 
few modes which are in \emph{avoided crossing}.  We find  the first-order 
correction to the frequency as the condition for the existence of non-trivial
solutions of (Eq.~\ref{eq:defsys-2}). 
\eqn{ \Bigl(-\wut+\wua+\frac{\dst \delta \omega_0}{\dst 2} \Bigr)
\Bigl(-\wut+\wub-\frac{\dst \delta \omega_0}{ \dst 2}\Bigr)
- {\cal W}_{1,ab} =0\,,\label{eq:finABo1}}
where
\eqn{{\cal W}_{1,ab} =\wuab\,\wuba\,.\label{eq:defW1ab}}
In the case of two degenerate modes, this system is equivalent to Eq.~59--60 in 
\dg. Conditions for non-trivial solutions give the first-order  frequency 
corrections in presence of near degeneracy
\eqn{\wut=\frac{\wua+\wub}{2}\pm\dst\sqrt{{\cal H}_{1,ab}}\,,\label{eq:solwuto1}}
with
\eqn{{\cal H}_{1,ab}= \Big(\frac{\wua-\wub}{2}+\frac{\delta \omega_0}{2}\Big)^2+
{\cal W}_{1,ab}\,.\label{eq:defH1ab}}
Note that, in case  of negligible  near degeneracy, that is, when
$4\omega^2_{1,ab}<<(\wua-\wub-\delta \omega_0)^2$, the non-degenerate $\wua$ and 
$\wub$ frequencies are retrieved.
\begin{figure}
   \hspace{-0.8cm}
   \includegraphics[width=9.9cm]{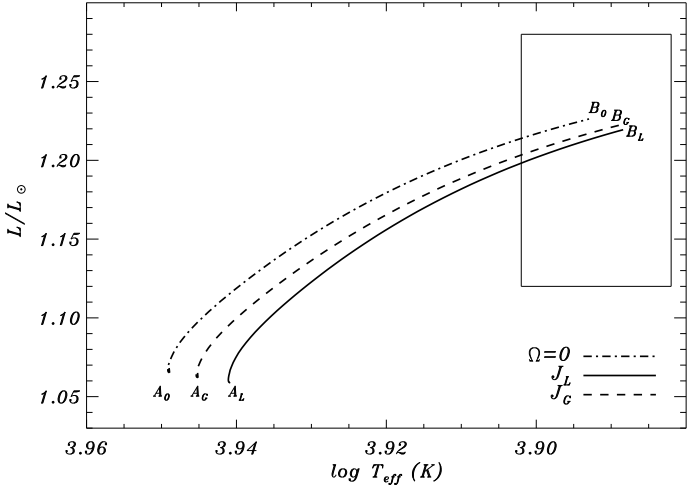}
   \caption{Evolutionary tracks of $1.8\,\msol$ models representative 
            of a typical \ds. The three tracks correspond to: a uniform 
	    rotating model (dashed line), a shellular rotating model 
	    (continuous line) and finally, a non-rotating model 
	    (dash-dotted line). The box represents typical observational 
	    errors for \dss.}   
   \label{fig:ComparTracksJGJL}
\end{figure}
The second-order near-degenerate frequency correction is again obtained as the 
condition for a non-trivial solution of Eq.20:
\eqn{\Big(\nu_a+\mu_a-\tilde \omega_2 
+\frac{\dst \delta \omega_0^2}{\dst 8 \bar \omega_0}\Big)\Big(\nu_b+\mu_b-\tilde \omega_2+
\frac{\dst \delta \omega_0^2}{\dst 8 \bar \omega_0}\Big)
-{\cal W}_{2,ab}=0\,.\label{eq:finABo2}}
where we have defined
\eqn{{\cal W}_{2,ab}=(\omega_{1,ab}+\mu_{ab})(\omega_{1,ab}+\mu_{ab})}
and
\eqn{\nu_b= \Big(-\wut+\wub-\frac{\dst \delta \omega_0}{ \dst 2}\Big)}
with $\tilde \omega_1$ given by Eq.26. Eq.~\ref{eq:solwuto1} provides two first 
order solutions, $\tilde \omega_1^{+}$ and $\tilde \omega_1^{-}$. Four possible 
solutions for $\wdt$ are thus obtained. Although all of them are mathematically 
valid, only two have a physical meaning. Each second-order solution 
$\wdt=\wdt^{+}$ and $\wdt^{-}$ is associated to one and only one first-order 
solution $\wut^{-}$ and $\wut^{+}$, respectively. We then obtain:
\eqn{\wdt= \Big(\frac{\nu_b+\nu_a}{2}+\frac{\mub+\mua}{2}+ 
\frac{\delta \omega_0^2}{8\bar \omega_0}\Big)\pm
\dst\sqrt{{\cal H}_{2,ab}^{(1)}}\,,
\label{eq:solsyscase12}}
where
\eqn{{\cal H}_{2,ab}^{(1)}=h(\nu,\mu)_{a,b}^2+{\cal W}_{2,ab}}
with
\eqn{h(\nu,\mu)_{a,b}=\frac{(\nu_a+\mu_a)-(\nu_b+\mu_b)}{2}\,.}
%
\begin{figure}
   \hspace{-1cm}
   \includegraphics[width=10cm]{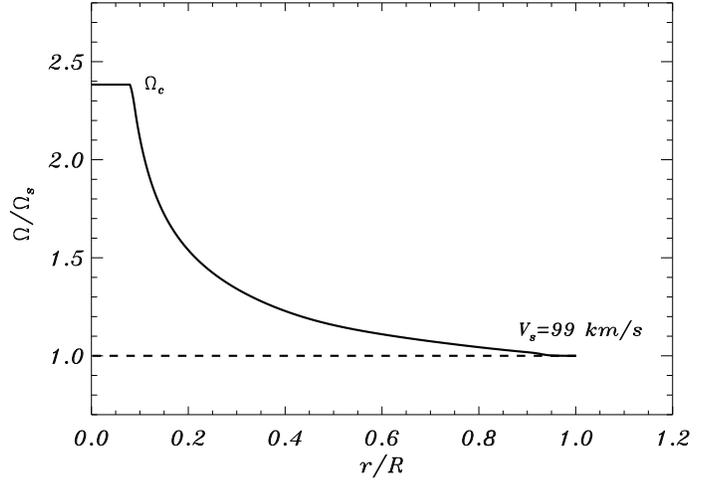}
   \caption{Rotation frequency through the star normalized to its 
            surface value. The rotation profiles correspond to $B_L$ 
	    model (curved line, shellular rotation) and $B_G$ model 
	    (horizontal line, uniform rotation) respectively. Models 
	    are built with the same surface rotation frequency 
	    $\Omega_{\mbox{s}}=9.98\,\muHz$. For more details see 
	    Sect.~\ref{sec:evolmodels}.}
   \label{fig:ComparProfils}
\end{figure}

\subsection{Second-order near degeneracy\label{ssec:SecondOrdDeg}}

In this case, $\delta \omega_0$ is $O(\Omega^2)$, $\wut=\omega_1$ and, $\omega_{1,jk}=0$ 
in Eq.~\ref{eq:Defmuj}. The quantity $\mu_j$ in Eq.~\ref{eq:Defmujk}  is then  
$\omega_{2j}$ as defined in Eq.~\ref{eq:w2m}. Only  modes with $\la=\lb\pm2$ are 
concerned. The analytic expression of second-order frequency corrections is obtained 
as the condition for existence of non-trivial solutions for 
\eqn{\sum_{j=a,b}\alpha_j\left[\mjk^{(2)}  
+ \frac{\delta \omega_0}{2}{\cal I}_{-1}\right]=0\,.\label{eq:finABo3}}
We then obtain
\eqn{\wdt= \Big(\frac{\mub+\mua}{2}\Big)\pm\dst\sqrt{{\cal H}_{2,ab}^{(2)}}\,,
\label{eq:solsyscase22}}
where
\eqn{{\cal H}_{2,ab}^{(2)}=\Big(\frac{\mu_b-\mu_a}{2}- \frac{\delta \omega_0}{2}\Big)^2
+ \muab \, \muba\,.\label{eq:defH2ab2}}
When near degeneracy is negligible, $4\muab\muba<<(\mu_a-\mu_b+\delta\omega_{0})^2$,
and the non-degenerate solutions $\mua= \omega_{2a}$ and $\mub= \omega_{2b}$
are respectively retrieved.

Our numerical computations suggest that for rapid enough rotations, every pair 
of modes having $\Delta\,\ell=\pm 2$ with radial orders $n$ and $n-1$ respectively, 
should present frequencies near enough to be coupled. For pairs with $\ell=0,2$, 
coupling is more significant for high radial order modes.
\begin{table}
  \begin{center}
    \caption{Characteristics of the three $1.8\,\msol$ models 
             considered in Fig.~\ref{fig:ComparTracksJGJL}. From left
	     to right, $\Omega_{s}$ represents the surface rotation frequency, 
	     $\teff$ the effective temperature, $X_{c}$ the central hydrogen 
	     fraction and $\bar{\rho}$, the mean density. Two evolutionary 
	     stages are considered: the zero age main sequence ($A$), and a 
	     main sequence age of 1050 Myr ($B$). A solar metallicity, a mixing 
	     length parameter of $\alpha_{ML}=1.614$ and, an overshooting 
	     parameter of $d_{\mbox{ov}}=0.2$ have been assumed.}
    \vspace{1em}
    \renewcommand{\arraystretch}{1.2}
    \begin{tabular}[h]{cccccc}
      \hline
        Model & $\Omega_{s}$  & $R$ & $\log\teff$ & $X_{c}$ &  $\bar{\rho}$ \\
               &  ($\muHz$) & ($R_\sun$)    &   (K)  &  & ($\mathrm{g}\,\mathrm{cm}^{-3}$)\\
      \hline
       $A_L$   & 15.485 & 1.824 & 3.870 & 0.726 & 7.308\\
       $A_G$   & 10.902 & 1.800 & 3.873 & 0.726 & 7.555\\
       $A_0$   & 0      & 1.795 & 3.876 & 0.726 & 7.927\\
       $B_L$   &  9.980 & 2.271 & 3.888 & 0.318 & 2.839\\
       $B_G$   &  9.757 & 2.275 & 3.888 & 0.315 & 2.827\\
       $B_0$   & 0      & 2.242 & 3.892 & 0.312 & 3.939\\
      \hline
      \end{tabular}
    \label{tab:ABmodeles}
  \end{center}
\end{table}

\section{Evolutionary models\label{sec:evolmodels}}
Equilibrium stellar models  have been computed with the evolution code \emph{CESAM}. 
Following the approach described in \cite{KipWeig90}, equilibrium models are 
constructed by modifying the stellar structure equations so as to include the 
spherical symmetric contribution of the centrifugal acceleration, by means of an 
\emph{effective gravity}.
\eqn{g_{\mathrm{eff}}=g-{\cal A}_{c}(r)\,,\label{eq:graveff}}
where $g$ is the local gravity component at a radial distance $r$ 
\eqn{g=G\,\frac{m(r)}{r^2}\,,\label{eq:gravite}}
with $m(r)$ the spherical mass at radius $r$, $G$ the gravitational constant  
and, ${\cal A}_{c}(r)$ corresponds to the centrifugal acceleration at radius 
$r$ :
\eqn{{\cal A}_{c}(r)=\frac{2}{3}\,r\,\Omega^2(r)\,.\label{eq:fcentrif}}
This spherically symmetric contribution of the rotation does not change the 
shape of the hydrostatic equilibrium equation. This is not fully representative 
of the structure of a rotating star. 

First, the non-spherical component of the deformation of the star is not 
considered. However, its effects are included through a perturbation in the 
oscillation equations (see previous sections). Second, rotationally-induced 
mixing and transport of angular momentum can significantly modify 
the rotation profile 
\citep{Talon97, MaederMeynet00, Heger00, Denissenkov00,Palacios03}. 
As these processes are not included in our models, we consider here two 
illustrative cases when prescribing $\Omega(r)$:

No mass loss is considered at any evolutionary stage, that is, the total 
angular momentum is assumed to be conserved. In this framework, the two 
assumptions for the transport of the angular momentum are: 1) either 
instantaneous transport of angular momentum  in the whole star (global 
conservation) which thus yields a uniform rotation, or, for sake of simplicity 
and illustrative purpose, 2) local conservation 
of the angular momentum (\emph{shellular} rotation). 
Formally, the local 
conservation of  angular momentum between two instants 
($t_2>t_1$) can be expressed as:
\eqn{\frac{2}{3}r_{1}^2(m)\,\Omega_{1}(m)=\frac{2}{3}r_{2}^2(m)\,\Omega_{2}(m)\,,}
where $m$ is the Lagrangian abscissa coordinate. Rotation in the convective 
core is considered to be rigid. 
\begin{figure}
  \begin{center}
   \includegraphics[width=8.75cm]{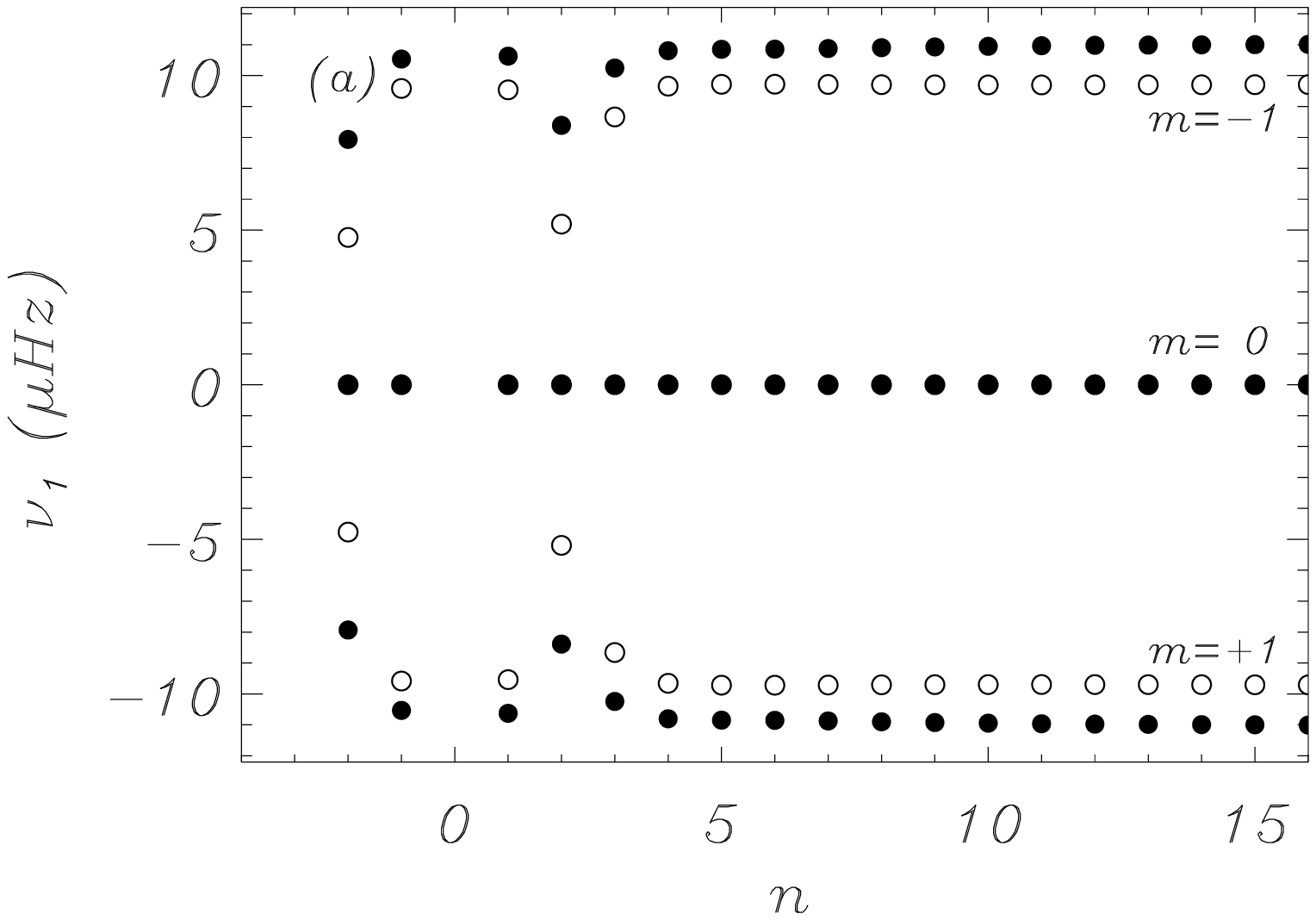}
   \includegraphics[width=8.75cm]{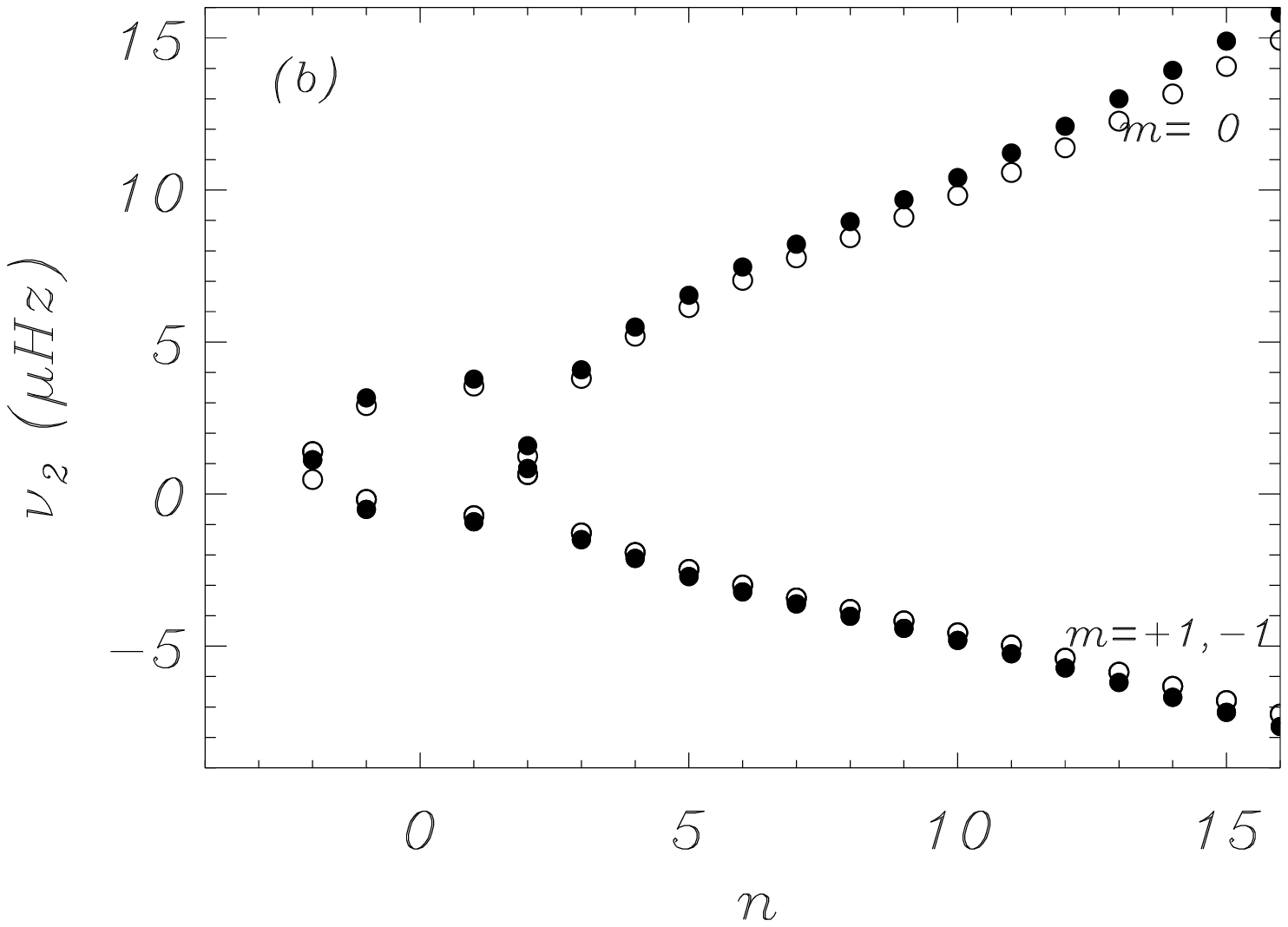}
   \includegraphics[width=8.75cm]{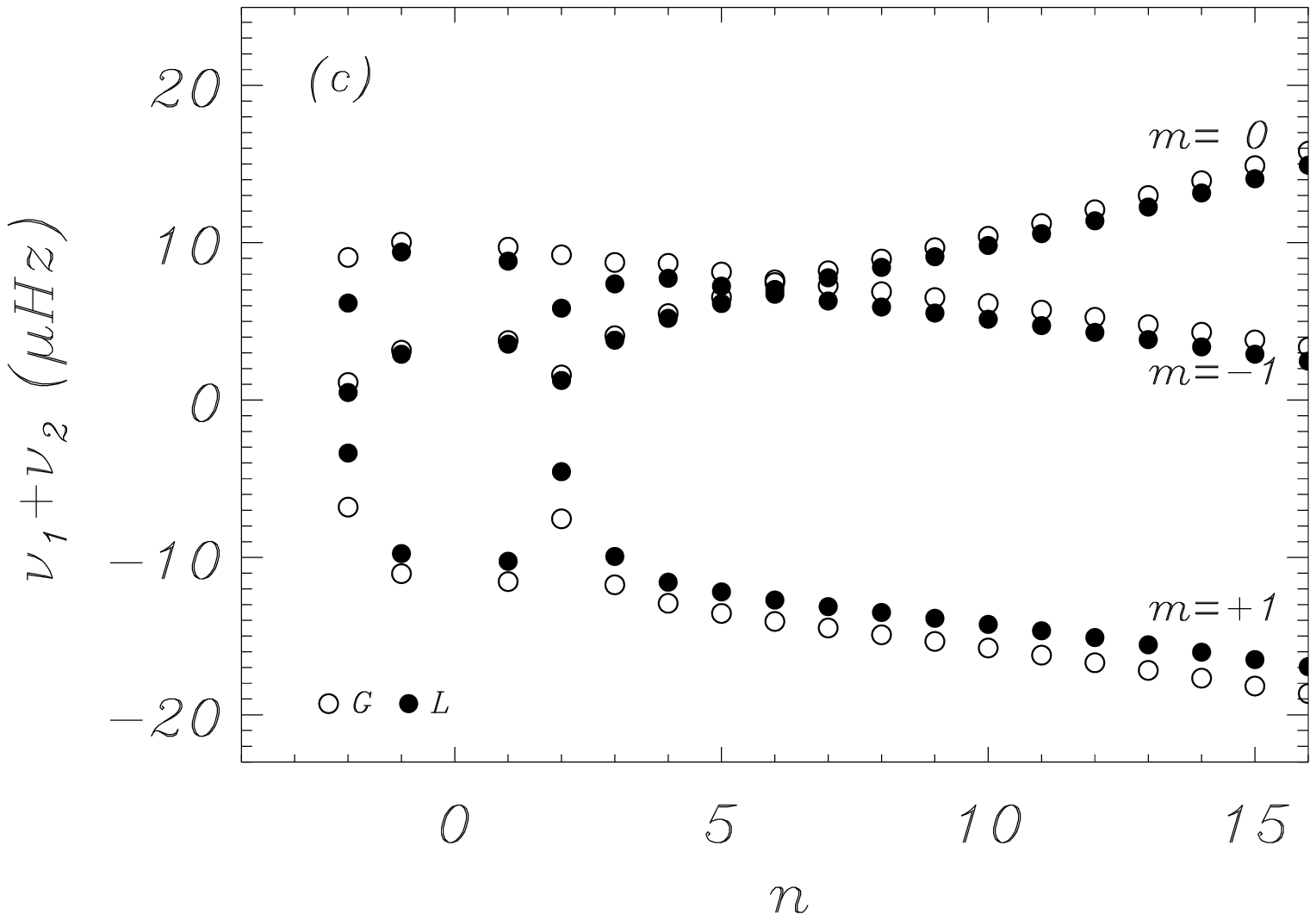}
   \caption{First (a), second (b) order frequency corrections 
            $\nu=\omega/2\pi$ for $\ell=1$ modes as a function of 
	    the radial order $n$. Finally, both contributions 
	    $\omega_1+\omega_2$ are displayed in panel (c). Empty 
	    and filled circles represent frequency corrections 
	    computed in the cases of uniform and shellular rotation 
	    respectively.}
   \label{fig:w1w2paper_n}
  \end{center}
\end{figure}
As shown by Eq.~\ref{eq:graveff}, gravity is modified by the radial term of 
the centrifugal acceleration ${\cal A}_c$ (see Eq.~\ref{eq:fcentrif}). 
In Fig.~\ref{fig:ComparAcelCentrif}, ${\cal A}_c$ is displayed as a function of
the radius for both cases (uniform and shellular rotation). The radial distance
$r=0.1\,R$ corresponds to a local maximum which
is the limit of the convective core. 
\begin{figure*}
  \hspace{-0.5cm}
  \includegraphics[width=8.75cm]{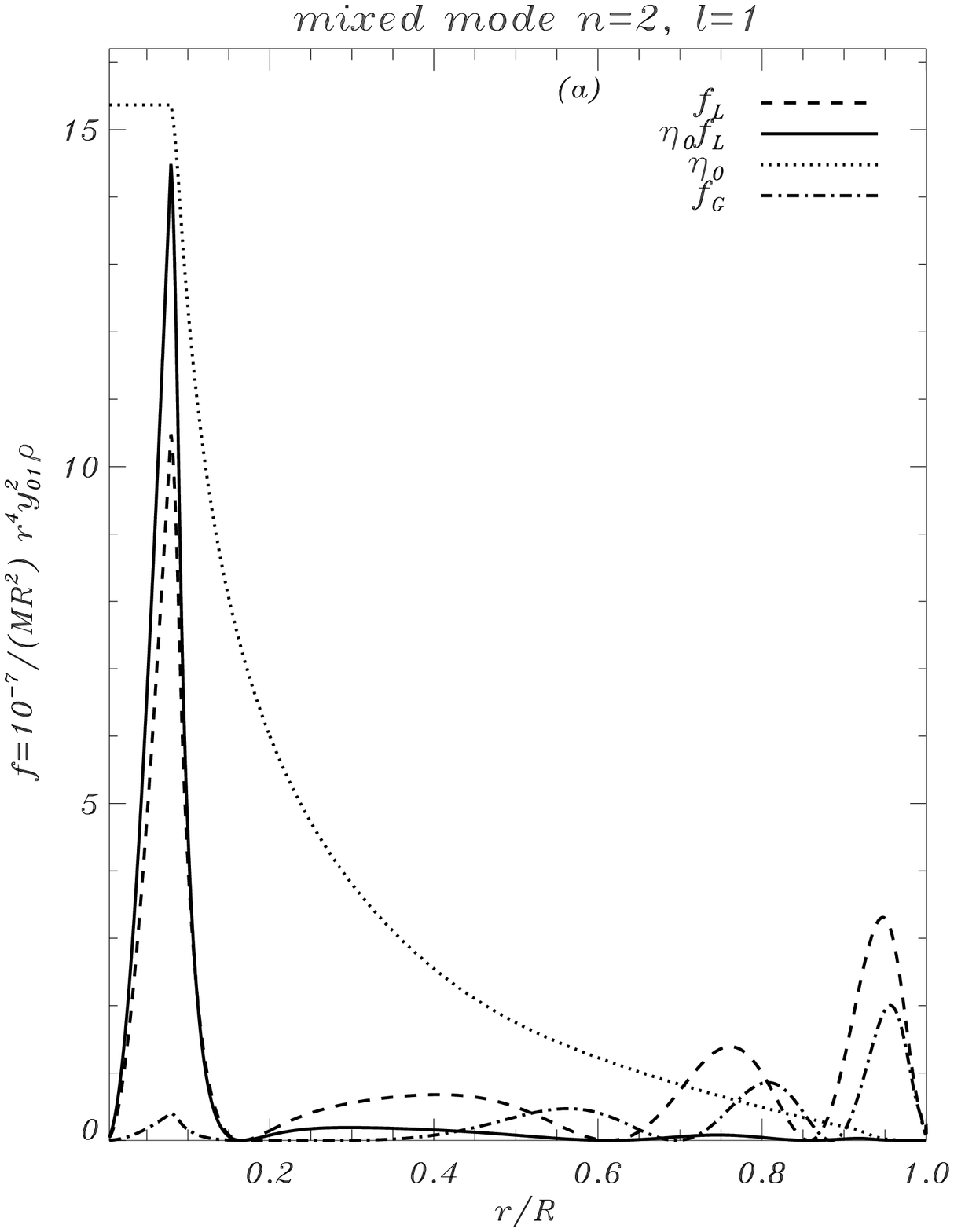}\hspace{-0.1cm}
  \includegraphics[width=8.75cm]{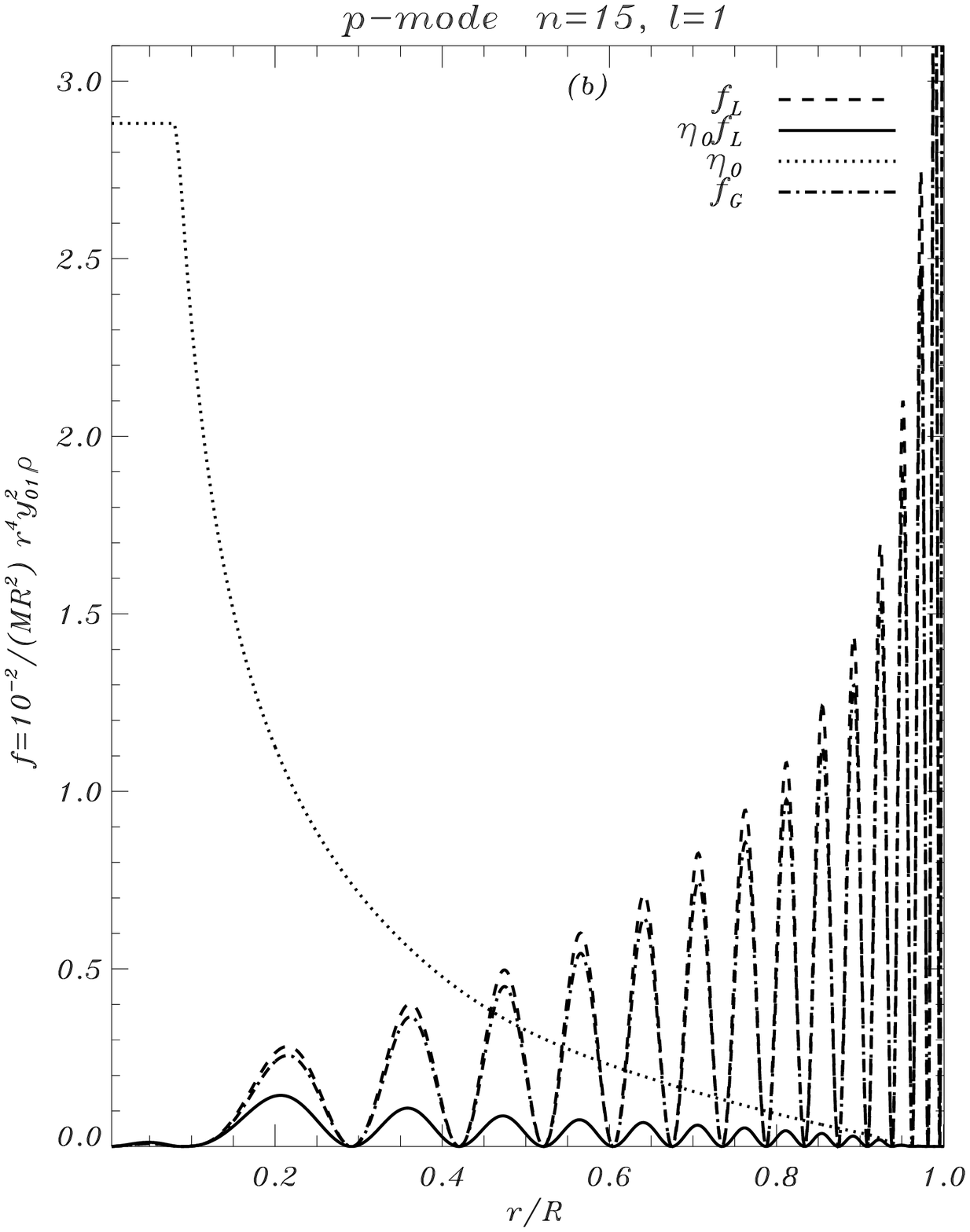}
  \caption{Weighted $\you$ functions versus the radial distance $r$ 
           (normalized to the radius of the star $R$). Solid and dashed 
	   lines represent the $f$ functions computed for the $B_L$ model.
	   Dash-dotted lines represent the $f$ function
	   (defined by Eq.~\ref{eq:fdef}) computed for the
	   $B_G$ model. Dotted lines represent the rotation profile 
	   given by the radial function $\eta_0(r)$ (the scale has been adapted
	   for the sake of clarity). Panel ($a$) shows $f$ computed
	   for the $n=2,\ell=1$ \emph{mixed} mode. Respectively, panel ($b$)
	   represents $f$ computed for the $n=15,\ell=1$ $p$ mode. In both plots,
	   the corresponding shellular rotation profile (same as 
	   Fig.~\ref{fig:ComparProfils}) is also displayed.}
\label{fig:eigenfuncPM}
\end{figure*}
This implies a modification of the local density, particularly in such regions.

In Fig.~\ref{fig:ComparTracksJGJL}, evolutionary tracks for a $1.8\,\msol$ model 
computed with  three different assumptions: uniform rotation ($J_G$), shellular 
rotation ($J_L$) and absence of rotation ($\Omega=0$). Two evolutionary stages 
are labeled: the zero age main sequence ($A_{}$, $A_{J}$ and $A_{L}$ models), 
and  on the main sequence ($B_{}$, $B_{J}$ and $B_{L}$ models). Fundamental 
stellar parameters of these models are given in Table~\ref{tab:ABmodeles}. During 
the main sequence evolution of the star, the rotation profile along the stellar 
radius mainly results from 1) the contraction of the core, and 2) the expansion 
of the outer layers. Then the local moment of inertia of contracting shells 
of a given elementary mass decreases. Therefore, the local rotation frequency
of those shells must increase in order to conserve the angular momentum
locally. Consequently, the rotation frequency toward the surface decreases 
during the evolution of the star (see Fig.~\ref{fig:ComparProfils}). This figure
shows a strong gradient of the rotation frequency toward the center: At a height 
of $r=0.2\,R$ the rotation rate is $\Omega=1.5\Omega_s$ with $\Omega_s$ the surface 
rotation rate. In our model, this means a rotation frequency around $14.97\,\muHz$ 
at $r=0.2\,R$, while a rotation frequency of $23.78\,\muHz$ is obtained at the 
limit of the convective core ($r\sim0.1\,R$). Oscillation modes having amplitude 
in these regions (typically $g$ and \emph{mixed} modes), are expected to be sensitive 
to shellular rotation.
\begin{figure*}
\begin{center}
\includegraphics[width=10cm]{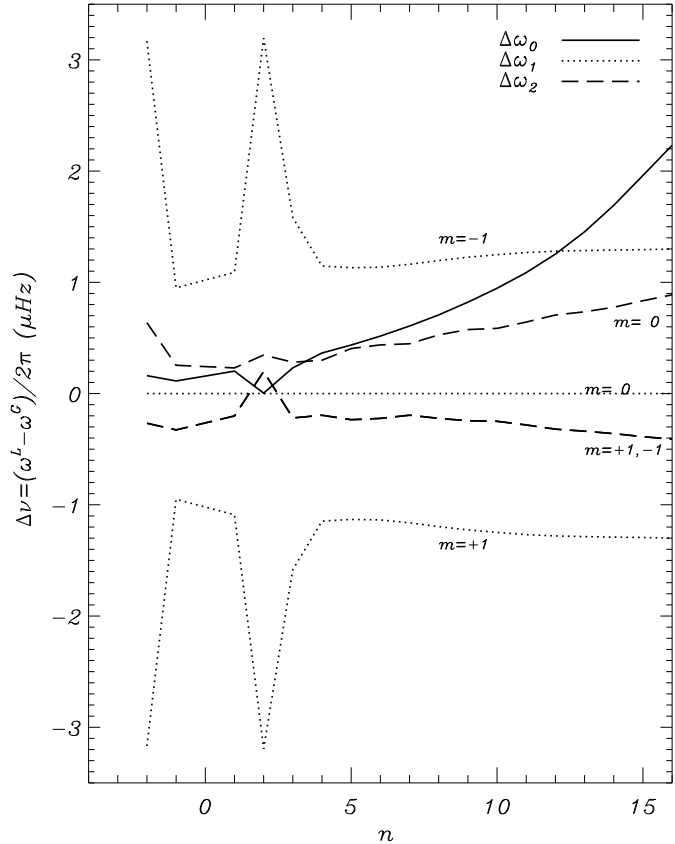} 
\caption{Differences between oscillation frequencies $\nu=\omega/2\pi$ 
         given in Fig.~\ref{fig:w1w2paper_n} as a function of their
	 radial order $n$. Such differences are displayed for the 
	 zeroth-order $\Delta\omega_0=\omega_0^{L}-\omega_0^{G}$ (represented 
	 by a continuous line); the first-order 
	 $\Delta\omega_1=\omega_1^{L}-\omega_1^{G}$ indicated by dashed 
	 lines, and finally for second-order corrections 
	 $\Delta\omega_2=\omega_2^{L}-\omega_2^{G}$ represented by dotted 
	 lines.}
\label{fig:w0w1w2dif_a}
\end{center}
\end{figure*}
In practice, observations provide the surface velocity (most often through the 
$\vsini$ parameter). In order to reproduce such conditions, models $B_L$ \& 
$B_G$ have been evolved so as to obtain the same rotation velocity at the surface. 
The first one ($B_L$) has been evolved considering the assumption of local 
conservation of the angular momentum. The second one ($B_G$) has been evolved 
under the assumption of uniform rotation (global conservation of the angular 
momentum).

\section{Effects of shellular rotation on oscillation frequencies
         \label{sec:effofDR}}

Oscillation frequencies of a rotating star are obtained from Eq.~\ref{eq:orders012}. 
For the sake of clarity and shortness, hereafter, the following nomenclature will 
be used: $\omega_0^{L}$ and $\omega_0^{G}$ for \emph{zeroth} order frequencies, 
$\omega_1^{L}$ and $\omega_1^{G}$ for first-order frequency corrections and finally, 
$\omega_2^{L}$ and $\omega_2^{G}$ will be used for second-order frequency corrections. 
The $L$ and $G$ superscripts follow the same nomenclature adopted for the models $B_L$ 
and $B_G$. Similarly, differences of frequency for a given mode (given $n,\ell, m$ 
subscripts) between two models will be denoted as 
$\Delta \omega_j= \omega_j^L- \omega_j^G$ for $j=0,1,2$ associated with zeroth-order 
frequencies, first- and second-order frequency corrections respectively.

It is found that the behaviour of frequency corrections does not significantly change 
for low $\ell$ degree values ($\ell\leq3$) when a uniform or a shellular rotation 
profile is assumed. We therefore focus on $\ell=1$ modes in this section. 

In the following sections, the analysis and discussion of the effect of shellular 
rotation on oscillation frequencies include very high-order $p$ modes. The reader 
should notice that such modes present a low probability of being observed (even from 
space) due to their expected small amplitudes. Even so, from a theoretical point of 
view, investigation of the frequencies of these modes -- i.e. with a well defined 
(asymptotic) behavior -- is found helpful for the global interpretation of the 
oscillation spectrum.

\subsection{Zeroth-order frequency corrections\label{sssec:eff0order}}

Perturbing and linearizing the hydrodynamical equations about a pseudo-rotating 
model yield an eigenvalue system: (see Eq.~\ref{ap.eq:sist_ordzero} in 
Appendix~\ref{ap:oscil-pseudo} for more details). Our numerical oscillation code 
is adapted to solve this system, with the boundary conditions specified in 
Eq.~\ref{ap.eq:condlim_center}--\ref{ap.eq:condlim_surf}, and including a radial 
shellular rotation profile (Eq.~\ref{eq:defeta0}). The resolution of this system 
yields the \emph{zeroth} order frequency $\omega_0$. The use of pseudo-rotating 
models described in Sect.~\ref{sec:evolmodels}, provides \emph{zeroth} order 
eigenfrequencies which include the effects of the spherically symmetric perturbation 
of pressure, density and gravitational potential. These perturbations are induced 
by the spherically symmetric component of the centrifugal acceleration (see \soufi). 
\begin{figure*}
  \begin{center}
  \includegraphics[width=10cm]{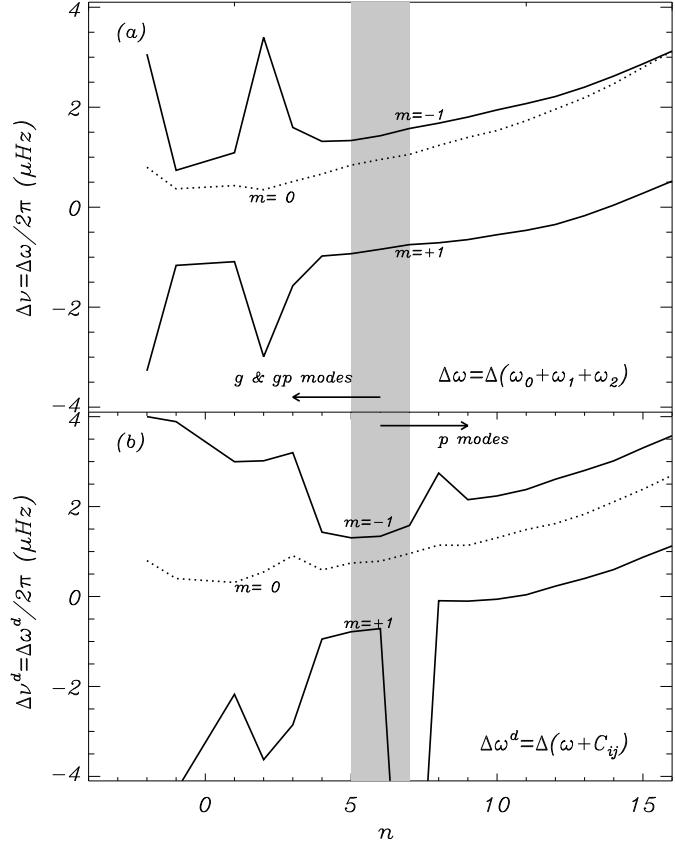}
  \caption{($a$) Same as Fig.~\ref{fig:w0w1w2dif_a} but 
            first- and second-order frequency corrections have 
            been included. Symmetric solid line branches represent from top 
	    to bottom, the differences for $m=-1$ and $m=+1$ mode frequencies 
	    respectively. For $m=0$ modes, differences are represented by a 
	    dotted line. Finally, panel ($b$) is equivalent to ($a$) but the 
            the correction for near degeneracy is included. 
	    In both panels, the shaded region represents an 
	    indicative frontier between the region of $g$ and \gp\ modes 
	    (left side) and $p$ modes (right side).}
\label{fig:w0w1w2dif_bc}
\end{center}
\end{figure*}
Eigenmodes of this system are still $m$ degenerate. 

Oscillation frequencies are thus computed for the $B_L$ and $B_G$ models described 
in Sect.~\ref{sec:evolmodels}. In order to quantify the effect of shellular rotation 
on zeroth-order frequencies, $\Delta\omega_0=\omega_0^{L}-\omega_0^{G}$ differences 
are calculated. These differences  are shown for  $\ell=1$ modes in 
Fig.~\ref{fig:w0w1w2dif_a} (solid line) as a function of the radial order $n$ of the 
mode. As can be seen, the effect of shellular rotation on \emph{zeroth}-order 
frequencies is very small at low frequencies and increases with the radial order $n$. 

For high radial order $p$ modes ($n \geq 12$), the differences $\Delta\omega_0$ are 
larger than differences from higher-order corrections $\Delta\omega_1,\Delta\omega_2$,  
(Sects.~\ref{ssec:PremOrdDeg}, \ref{ssec:SecondOrdDeg}) and hence become the most 
sensitive to the effect of shellular rotation. 

Numerical tests show that the differences $\Delta\omega_0$ mainly come from the 
differences in the pseudo rotating models, $B_L, B_G$; the contribution of the rotation 
profile (either uniform or shellular ) in the oscillation equations is negligible.

As an order of magnitude, for $\ell=0, n\sim 10-15$ modes, we consider the slope, 
$\nu_a$  ($\nu = \omega/2\pi)$, of the variation of the differences  
$(\nu_{0,n+1,\ell=0,0}- \nu_{0, n,\ell=0,0})$ with the radial order $n$ (in the asymptotic 
approximation). We numerically find that 
$\Delta \nu_a = (\nu^L_a -\nu^R_a) \sim 0.175\,\muHz$ for $\nu_a \sim 47\,\muHz$ i.e. 
a relative difference $\Delta \nu_a/\nu_a \sim 3.7 ~10^{-3}$. This difference can be 
accounted for by the fact that our $B_L$ and $B_G$ models have slightly different radii 
$\Delta\,R/R=(R^{L}-R^{G})/R^L = -1.76 ~10^{-3}$ (see Table~\ref{tab:ABmodeles}). 
This induces a contribution to $\Delta \nu_a$ which can be quantified as: 
\eqn{
\frac{\Delta\nu_a}{\nu_a}=-\frac{3}{2}\frac{\Delta\!R}{R}\sim 2.6 ~10^{-3}\,.
}
Hence in the present case, quite large differences, $\Delta\omega_0$, at high frequency, 
come from structural properties of the two models through differences in their structural 
variables (pressure, density perturbations... ) rather than from differences in the 
wave propagation properties which appear as additional extra terms in the oscillation 
equations (see Appendix~\ref{ap:oscil-pseudo}).

\subsection{First-order frequency corrections\label{sssec:eff1order}}

In Fig.~\ref{fig:w1w2paper_n}a, the first-order frequency  corrections $\omega_1$ 
(Eq.~\ref{eq:w1m}) are represented as a function of the radial order $n$. Empty circles 
represent the frequency corrections obtained for the $B_G$ model, i.e. in the case 
of uniform rotation. Respectively, filled circles are those obtained for the $B_L$ 
model, that is, in the case of shellular rotation (i.e. assuming local conservation 
of the angular momentum).
 
Figure~\ref{fig:w1w2paper_n}a can be seen as composed of two domains: a first one 
($n\leq5$) containing $g$~modes and $mixed$ modes (hereafter \gp\ modes region), 
and a $p\,$-mode region ($n>5$). As expected, modes in \gp\ region are more affected 
by shellular rotation than those in the $p\,$-mode region. 

In Fig.~\ref{fig:w0w1w2dif_a}, the frequency correction differences, 
$\Delta\omega_1=\omega_1^{L}-\omega_1^{G}$, are represented as a function of their 
radial order $n$ for $\ell=1$ modes. These differences assess the effect of assuming 
a shellular rotation instead of a uniform rotation. In the region of \gp\ modes, the 
effect of shellular rotation can reach up to $3\,\muHz$ for a few modes. 

For high radial order $n$, differences $\Delta \omega_1$, although smaller than for 
\gp\ modes, are significant and remain quite stable around $1\,\muHz$.

These results can be explained by analysing the contribution of shellular rotation 
to $\omega_1$ corrections (Eqs.~\ref{eq:w1m}, \ref{eq:defCl} \& \ref{eq:defJ0}), 
which arises from:
\begin{itemize}

  \item[1)] the effect of $\Omega(r)$ on the zeroth-order eigenfunctions $\you$ and 
            $\zo$ (corresponding respectively to the radial and horizontal displacements) 
	    appearing in $C_L$ and $J_0$ terms. Those eigenfunctions are obtained by 
	    solving the eigenvalue system taking into account a shellular rotation, 
	    $\Omega(r)$, given in Appendix~\ref{ap:oscil-pseudo} 
	    (Eqs.~\ref{ap.eq:defy01}--\ref{ap.eq:defy04}). Importance of this effect 
	    can be measured with the weighted function $f$ as defined by
            \eqn{f(r) = \rho\,r^4 y^2_{01}\,, \label{eq:fdef}}
            where $\you$ represents the radial displacement eigenfunction.

  \item[2)] the explicit contribution of the rotation profile $\eta_0$ in $J_0$ 
            (Eq.~\ref{eq:defJ0}). This latter term vanishes in case of uniform rotation. 
	    For high-frequency $p$ modes, the horizontal component of the displacement 
	    is negligible ($|y_{01}| >> z_0$) and one has $J_0 \sim <\eta_0 >$ where 
	    $<\cdot>$ is a weighted average defined as:
            \eqn{<\cdot> \sim \dst 
            \frac{\int_0^R ~\rho\,r^4  y^2_{01}~ \cdot~\dr}{\int_0^R~\rho\,r^4 y^2_{01}~\dr}\,.		 
                \label{eq:defaverage}} 
\end{itemize}
In Fig.~\ref{fig:eigenfuncPM}, the weighted function $f(r) = \rho\,r^4 y^2_{01}$ and 
$\eta_0(r) f(r)$ are then represented as function of the radial distance $r$ for a 
high radial order $p$ mode (Fig.~\ref{fig:eigenfuncPM}b) and for comparison in 
Fig.~\ref{fig:eigenfuncPM}$a$ for a \emph{mixed} mode. 

For \gp\ modes, the  eigenfunctions (as shown by $f$ functions in 
Fig.~\ref{fig:w1w2paper_n}a) present an inner maximum near the core $r/R\sim[0,0.2]$. 
As the kinetic energy of these modes are large in core-close regions, large 
differences between the uniform and the shellular rotation cases i.e. in 
$\Delta \omega_1$ are reasonably expected.

For high radial order $p$ modes, \emph{implicit} effects of shellular rotation on 
eigenfunctions $f_L$ are represented by dashed lines and can be compared with $f_G$ 
for a uniform rotation. Maximum amplitudes of such modes are located near the stellar 
surface, the eigenmodes do not differ much between the uniform and the shellular cases. 
The main contribution to the differences $\Delta\omega_1=1\,\muHz$ comes from the 
large increase of the rotation rate toward the center which is accounted for by 
$\eta_0$ in $J_0$. Hence, negative values of $J_0$ explain that $\omega_1$ absolute 
values obtained in the case of a shellular rotation are lower than those obtained when 
a uniform rotation is assumed as seen in Fig.~\ref{fig:w1w2paper_n}a.

\subsection{Second-order frequency corrections\label{ssec:eff2order}}

Second-order frequency corrections $\omega_2$ are found smaller than the first-order 
frequency corrections except at large frequency. They are presented in 
Fig.~\ref{fig:w1w2paper_n}b for $\ell=1$ modes as well as the sums ($\omega_1+\omega_2$) 
as a function of their radial order $n$. At low frequencies ($n \le 5$), $\Delta\omega_2$ 
is comparable with the effect on second-order corrections ($\Delta\omega_0$). 

In the $p$-modes region, investigations of the individual terms in  
Eq.~\ref{eq:w2tot_saio} reveal that $X_2$ and $Y_2$ are clearly dominant in that 
region. For given $\ell=1$ centroid modes, $\omega_2$ can be written as:
\eqn{\omega_2(\ell=1,m=0)=
\frac{\rotbar^2}{\omega_0}\left[X_1^T+X_1^I+X_2\right]\,.
      \label{eq:omega2l1m0}}
In this expression, $X_2$ is found to be dominant mainly due to its $\wo$ dependence 
through the ${\cal I}_c$ term (see Eq.~\ref{ap.eq:defIc_1}). Neglecting then $X_1^T$ 
and $X_1^I$, and considering $X_2(\ell=1,m=0)$, Eq.~\ref{eq:omega2l1m0} reduces to:
\eqn{\omega_2(\ell=1,m=0)\sim\frac{2}{5}
     \frac{\rotbar^2}{\omega_0}{\cal J}_c\,.}
For $m=\pm1$ modes, $Y_2$ behaves like $X_2$ but with opposite sign. Similarly to $X_2$, 
the presence of $C\wod$ in ${\cal J}_c$ makes $Y_2$ dominant face to $Y_1$. Thus 
the shape of the $m=\pm1$ branches shown in Fig.~\ref{fig:w1w2paper_n}b can  be explained 
by the \emph{competition} between $X_2$ (positive) and $Y_2$ (negative), yielding:
\eqn{\omega_2(\ell=1,m\neq0)\sim-\frac{1}{5}
     \frac{\rotbar^2} {\omega_0}  {\cal J}_c\,.   }
More generally, for high radial order ($\ell\neq 0,m\neq 0$) modes, one has (see also DG92):
\eqn{\omega_{2,m} \sim  
\frac{\left(\Lambda-3\,m^2\right)}{4\Lambda-3} \frac{\rotbar^2} {\omega_0}  {\cal J}_c \,.  
       \label{eq:depfromasym}}
Unlike the first-order frequency corrections, the second-order  $m\neq 0$ branches 
are not symmetric with respect to $m=0$ modes and the asymmetry for a given $\ell$ 
can be written as:
\eqn{ \omega_{2,m}+\omega_{2,-m}-2 \omega_{2,0}=   
 -\frac{6 m^2}{4\Lambda-3}  \frac{\rotbar^2} {\omega_0}  {\cal J}_c\,.}
For high radial order $p$ modes, ${\cal J}_c$ is given to a good approximation by 
\eqn{{\cal I}_c \sim \sigma^2_0 <{\cal S}_2 >\,,}
where $\sigma_0^2 = \omega_0^2/(GM/R^3)$ $<{\cal S}_2>$ is a  weighted average of 
perturbations of structure (see Appendix~\ref{ap:saioform}, Eq.~\ref{ap.eq:defIavecS2}). 
Hence,
\eqn{\omega_2(\ell=1,m=0)\sim\frac{2}{5}  \frac{\rotbar^2}{GM/R^3} \omega_0 <{\cal S}_2 >\,,}
and the increase of $\omega_2$ with $n$ clearly seen in Fig.~\ref{fig:w1w2paper_n}b 
arises from its $\omega_0$ dependence.
\begin{figure}
  \begin{center}

  \includegraphics[width=8.75cm]{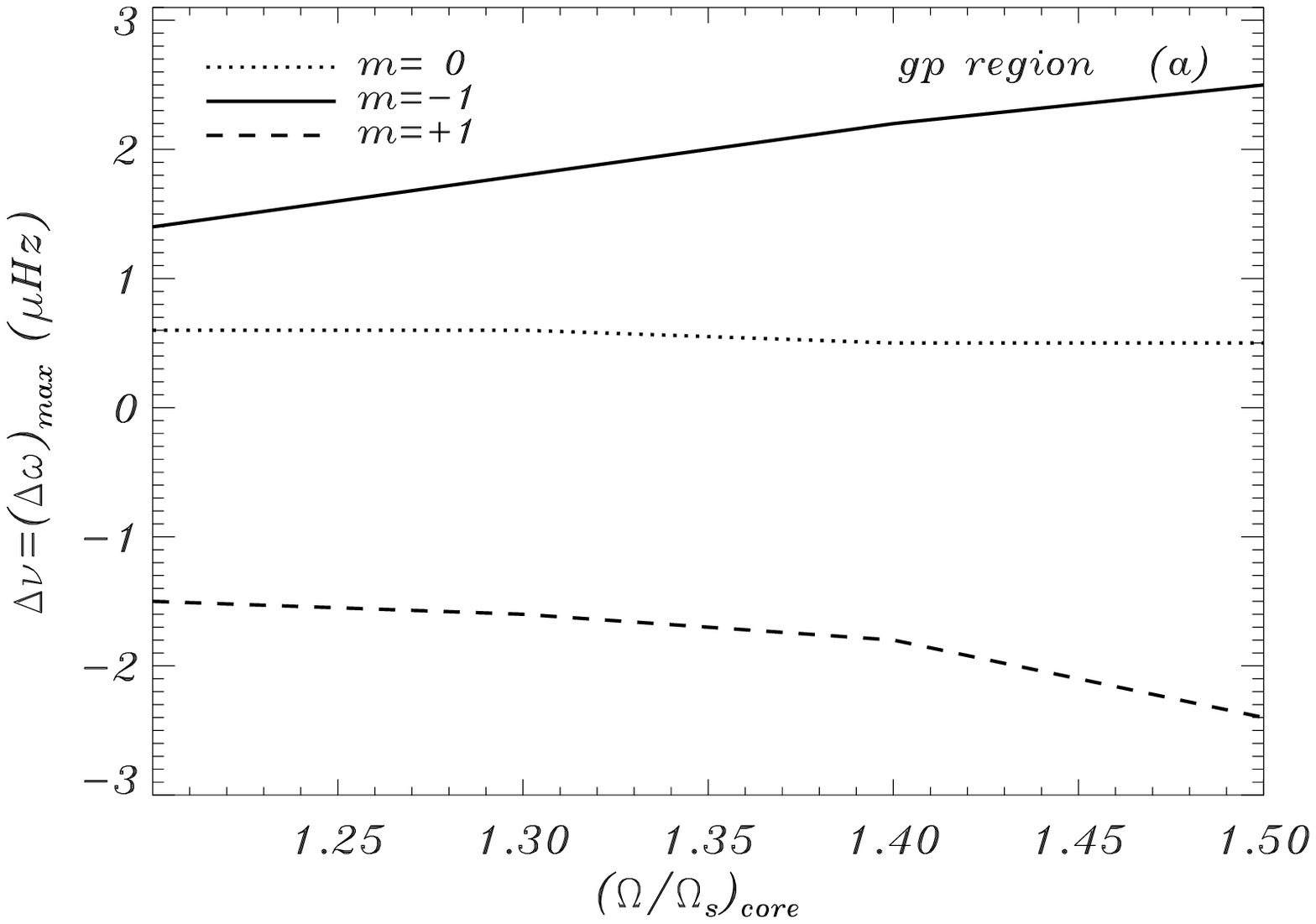} 
  \includegraphics[width=8.75cm]{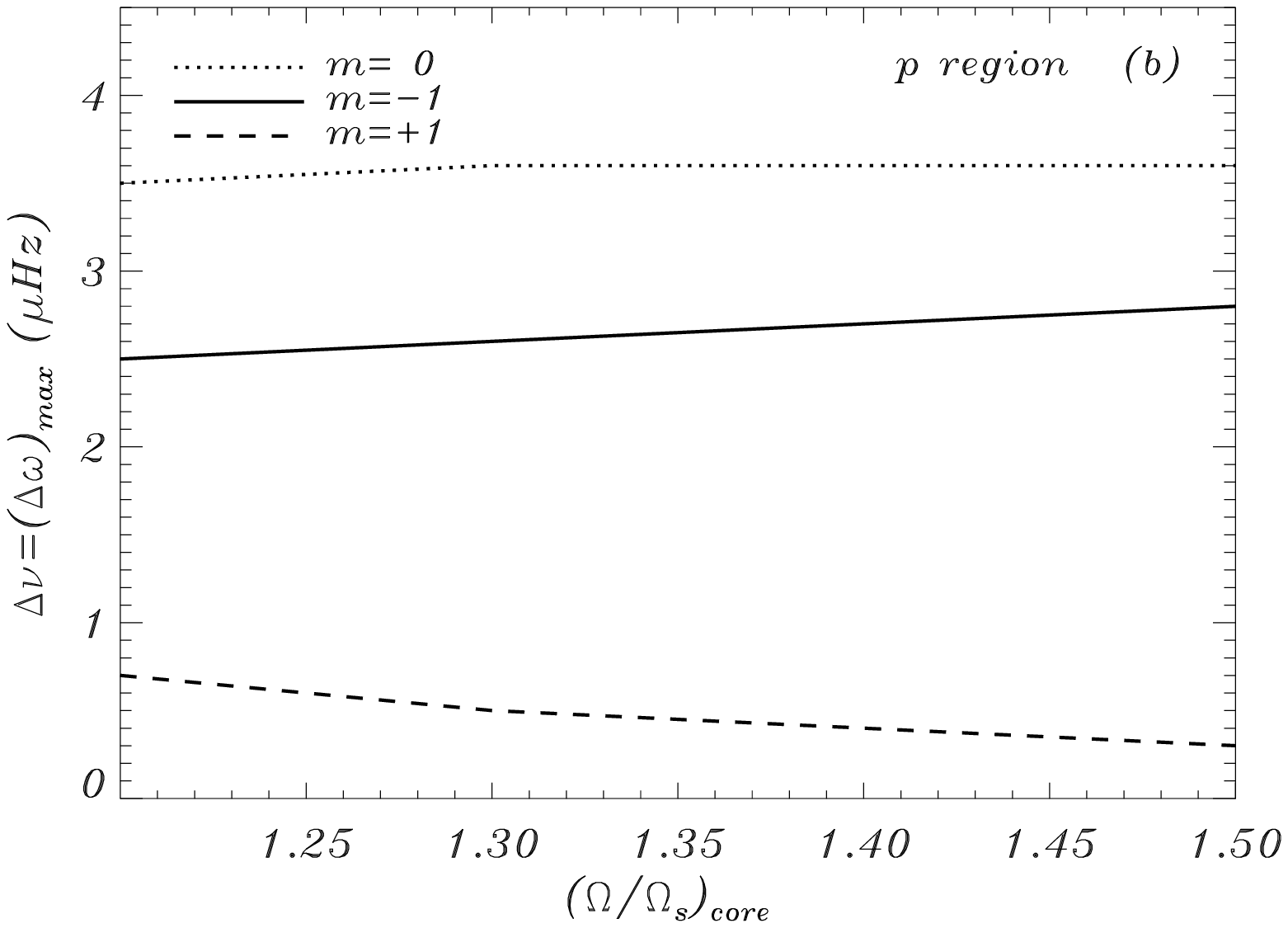}
  \caption{Variation of $\Delta\nu=\Delta\omega/2\pi$ with 
           $\Delta\omega=\Delta\omega_0+\Delta\omega_1+\Delta\omega_2$
	   as a function of the rotation rate of the uniformly rotating 
           convective core normalized to the surface rotation rate 
           $\Omega/\Omega_s=1.2$. Effect of varying the rotation rate is 
           shown for modes  in \gp- and $p$-modes regions. The plots focuse 
           on $\ell=1,m=0,\pm1$ modes for which $\Delta\omega$ gets a maximum
	   value (see Fig.~\ref{fig:w0w1w2dif_bc}a): $n=3$ modes in the 
           \gp\ domain (panel~($a$)),and $n=16$ modes in the  $p\,$-modes 
           region (panel~$b$).}
  \label{fig:maxdif_eta}
\end{center}
\end{figure}
In the \gp-modes region, in contrast to pure $p$ modes, $X_1, Y_1$ (more 
specifically $X^T_1, Y^P_1$) dominate over the structure rotational distortion 
effects $X_2, Y_2$. In that case, both effects 1) and 2) mentioned above contribute 
in an intricate way.

The combination of both first- and second-order frequency corrections in 
Fig.~\ref{fig:w1w2paper_n} is then found to be dominated by the $\omega_2$ behavior 
at high frequency and as it was the case of first-order corrections, $\omega_2^{L}$ 
present absolute values higher than $\omega_2^{G}$. At low frequency (\gp-modes region), 
a complicated behaviour results from both contributions ($\omega_1+\omega_2$). 

Figure~\ref{fig:w0w1w2dif_a} shows the frequency differences 
$\Delta \omega_2= \omega^L_2-\omega^G_2$. In general $\Delta\omega_2<\Delta\omega_1$ 
over the whole range of radial order considered. Contrary to the first-order results, 
the effect of shellular rotation on second-order frequency corrections 
($\Delta\omega_2$) is larger for $p$ modes than for \gp modes. 

For high radial order $p$ modes, the second-order frequency differences arise from:
\eqn{\Delta\omega_2\sim \frac{\rotbar^2}{GM/R^3}\Big[
     \Delta(\omega_0 <{\cal S}_2> )\Big]\,,
      \label{eq:2ordeffapproxX2Y2}}
with
\eqn{\Delta(\omega_0  <{\cal S}_2> )=\omega_0^{L}  <{\cal S}_2>^{L}-\,\omega_0^{G} <{\cal S}_2> ^{G}.}
As shown in Fig.~\ref{fig:w0w1w2dif_a}, in the $p\,$-mode region, $\Delta\omega_0$ 
increases with the radial order $n$. The effect of this increase is lowered in  
$\Delta\omega_2$ by the change of $<{\cal S}_2>$ which itself is mainly due to the 
derivatives of $\Omega(r)$ (i.e. $b_2, db_2/dr$).

For \gp\ modes, the difference in $\omega_0$ between both models is small 
(Fig.~\ref{fig:w0w1w2dif_a}) and its contribution to $\Delta\omega_2$ (Eq.44) is 
less important. In this case, second-order effects of centrifugal and Coriolis 
forces on the stellar structure are dominant through terms in ${\cal J}_c$ 
proportional to:
\eqn{r\deriv{u_2}{r},~r u_2,~r\bd\,.}
This is shown in equations describing the behaviour of the pressure and density 
distributions in presence of radial shellular rotation (Eqs.~74--80 in \dg). 
Moreover, second-order effects of shellular rotation on the non-spherically 
symmetric component of the gravitational potential $\phi_{22}$ are found to be 
negligible and $u_2\sim (1+\eta_2)$ in Eq.~\ref{ap.eq:defIc_1}. Therefore, 
$\Delta\omega_2$ differences are dominated by the effect of $\Omega(r)$ ($\eta_0$) and 
of its derivatives ($b_2$,...).

As it was the case of first-order corrections, $\omega_2^{L}$ present absolute 
values higher than $\omega_2^{G}$.

\subsection{Full non-degenerate frequencies}

Finally, the total effect of differential rotation up to second order on the frequency 
$\Delta\omega$ is displayed in Fig.~\ref{fig:w0w1w2dif_bc}a.

The largest effects of shellular rotation on \gp\ modes (up to $\sim3.4\,\muHz$) arise 
for $m=\pm1$ modes and are dominated by $\Delta\omega_1$. In this region, the effects 
on \emph{zeroth} order frequencies are much smaller. In contrast, for the $p$-modes 
region $\Delta\omega$ are dominated by $\Delta\omega_0$ increasing with the radial order 
$n$ (see Sect.~\ref{sssec:eff0order}). Numerically, we find differences up to 
$0.5\,\muHz$ for $m=+1$ modes. For $m=-1$ modes, effects up to $3\,\muHz$ are seen to 
be of about the same order than those found for $m=0$ modes. 

As already mentioned, the efficiency of the transport of angular momentum is modified 
by rotational mixing during the evolution of the star. The effect of shellular rotation 
on oscillation frequencies is expected to be reduced (in preparation), due to the 
\emph{smoothing} of the variation of $\Omega/\Omega_s$ at the edge of the convective 
core (see Fig.~\ref{fig:ComparProfils}). In order to simulate such a situation, 
theoretical oscillation spectra for the model $B_L$ have been computed with core rotation 
rates in the range of $\Omega/\Omega_s=[1.5,1.2]$. As expected, the major influence of 
such variations is found for \gp\ modes. The lower the rotation rate in the core, 
the lower the effect of shellular rotation in this region 
(see Fig.~\ref{fig:maxdif_eta}a). For the minimum rotation rate considered, 
$|\Delta\omega|\sim1.4\,\muHz$ for $m=-1$ modes, and $|\Delta\omega|\sim1.5\,\muHz$ for 
$m=+1$ modes is found. For $p$ modes (see Fig.~\ref{fig:maxdif_eta}b), differences 
remain close to those obtained for $\Omega/\Omega_s=2.5$ (used along the present work).

\begin{figure*}
  \hspace{-0.5cm}
  \includegraphics[width=9cm]{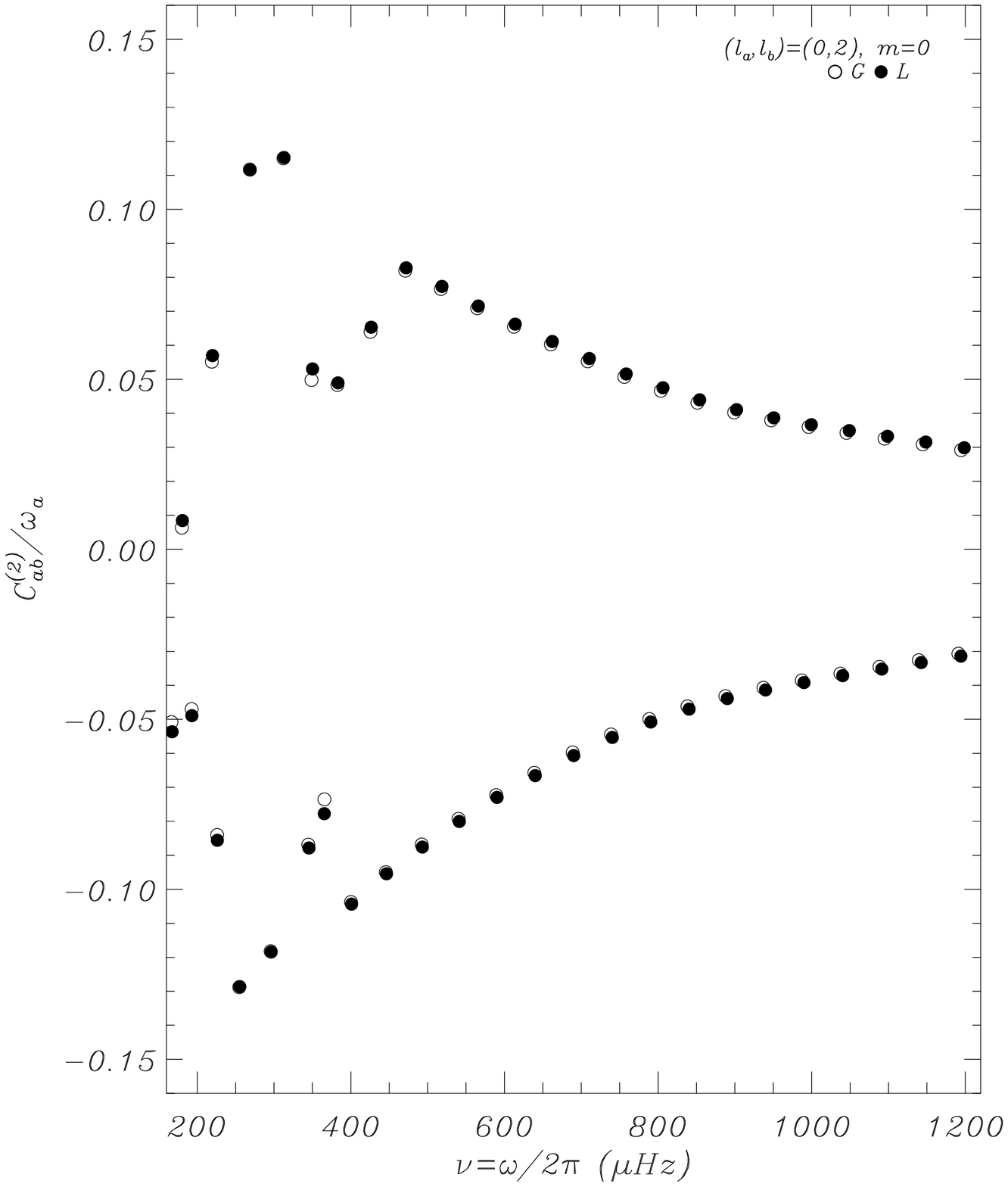}  
  \includegraphics[width=9cm]{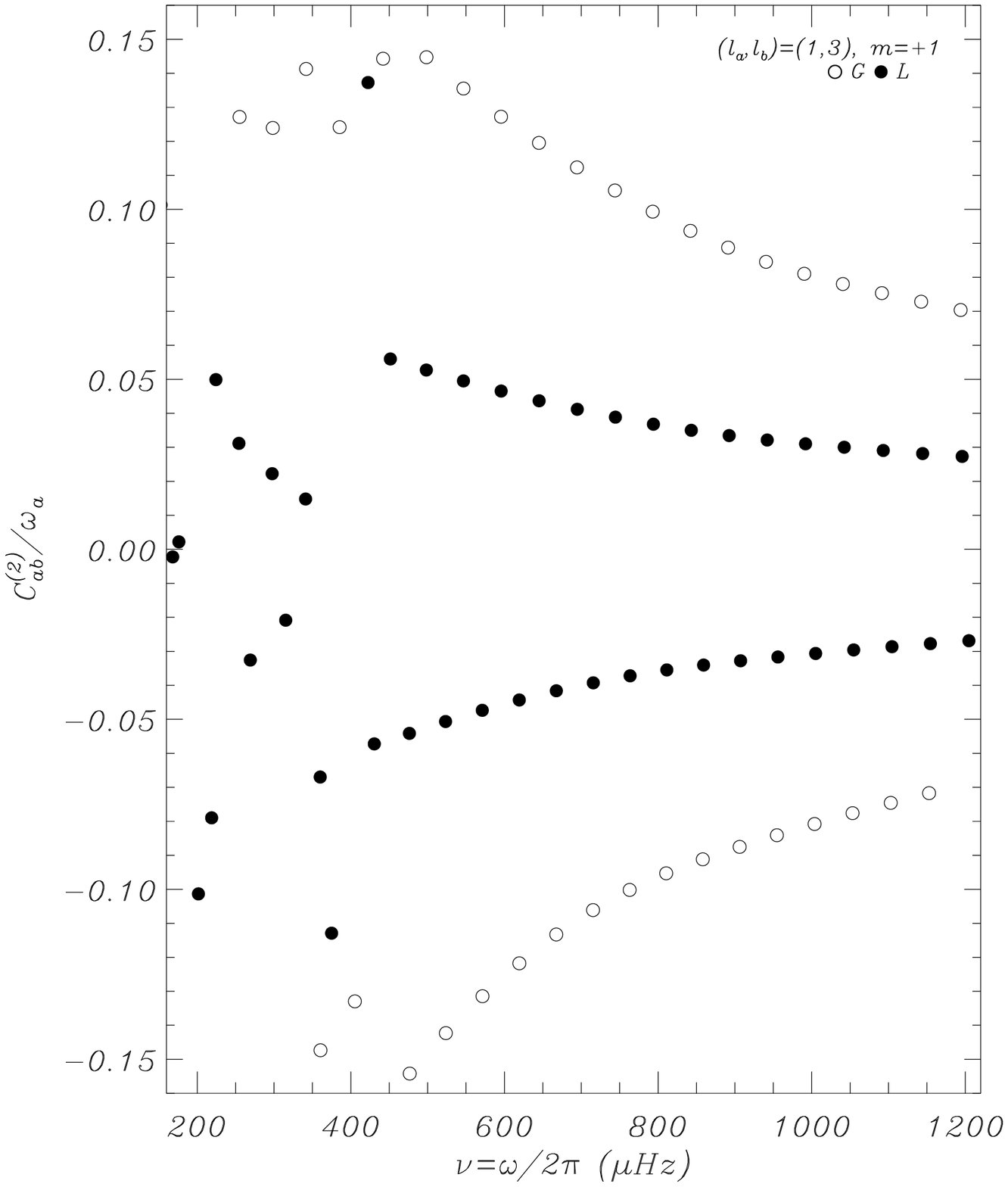}  
  \caption{Effect of the second-order near-degenerate coefficients 
           $C_{ab}^{(2)}/\omega_a$  and $C_{ab}^{(2)}/\omega_b$ as 
           a function of the frequency ($\nu=\omega/2\pi$) for 
           degenerate pairs $(\ell_1,\ell_2)=(0,2)$ (left panel) 
           and $(\ell_1,\ell_2)=(1,3)$ (right panel). Empty and 
           filled circles represent degenerate coefficients in the 
           cases of uniform and shellular rotation respectively.}
   \label{fig:cij2-w}
\end{figure*}
\subsection{Correcting for near degeneracy\label{ssec:effNearDeg}}

Near degeneracy is taken into account as described in Sect.~\ref{sec:neardeg}. 
The oscillation frequency of near-degenerate modes are computed according to 
Eq.~\ref{eq:defw}, \ref{eq:solwuto1} \& \ref{eq:solsyscase12} or 
Eq.~\ref{eq:defw} \& \ref{eq:solsyscase22}. 
In Fig.~\ref{fig:w0w1w2dif_bc}b, $\omega^{L}$ and $\omega^{G}$ are represented for  
$(\ell_a,\ell_b)=(1,3)$ degenerate pairs. Near degeneracy is considered here for 
modes with $|\sigma_a-\sigma_b|\lesssim\Omega_s/(GM/R^3)^{1/2} \sim 0.25$ 
($ i.e. |\nu_a-\nu_b| \sim 10\,\muHz here$) corresponding to the surface rotation 
frequency of $B_G$ and $B_L$ models (see Table~\ref{tab:ABmodeles}).
As can be noticed, for both models, modes in the \gp\ region clearly are the 
most affected by near degeneracy, with a few  $m=\pm1$ near-degenerate modes within 
the range of $n\lesssim8$. This result was expected since such modes are likely 
closer to each other than $p$ modes. In order to quantify how the oscillation 
frequencies are modified by near degeneracy, we define the coefficients $C^{(1)}_{ab}$ 
and  $C^{(2)}_{ab}$ composed for first- and second-order contributions, where $ab$ 
subscripts represent the two modes affected by near degeneracy. According to the 
definitions Eq.~\ref{eq:omegabar}--\ref{eq:defxi0}, "+" sign is for mode a and "-" 
sign is for mode b. In the case of near-degenerate modes with same degree $\ell$, 
the first- and second-order coefficients are respectively defined as 
\eqn{  C^{(1)}_{ab}\equiv \wut-\wu \pm \frac{\Delta \omega_0}{2},
     ~~C^{(2)}_{ab}\equiv \wdt-\omega_2 - \frac{\Delta \omega_0^2}{8\omega_0}\,. 
     \label{eq:defCij_12a}}
In the case of  two degenerate modes with different degrees $\ell_a=\ell_{b}\pm2$, 
the first-order coefficient $C^{(1)}_{ab}=0$ (see Sect.~\ref{sec:neardeg}) and 
\eqn{C^{(2)}_{ab}\equiv \wdt-\omega_2\pm\frac{\Delta \omega_0}{2}
\label{eq:defCij_12b}}
The full expression for the adiabatic oscillation frequencies can be written as:
\begin{equation*}
  \omega=
     \begin{cases}
            \omega_0+\big[\omega_1+C^{(1)}_{ab}\big]
	           +\big[\omega_2+C^{(2)}_{ab}\big]     
	            &\text{if $\ell_a=\ell_b$}\\
            \omega_0+\omega_1
	            +\big[\omega_2+C^{(2)}_{ab}\big]    
	            &\text{if $\ell_a=\ell_b\pm2$}
     \end{cases}\label{eq:freqtot_coup}
\end{equation*}

In Fig.~\ref{fig:cij2-w}, for degenerate pairs $(\ell_a,\ell_b)=(0,2)$ (left panel) 
and for $(\ell_a,\ell_b)=(1,3)$ (right panel), the effect of near degeneracy is 
assessed by $C_{ab}^{(2)}/\hat \omega_0$ as a function of the frequency for the cases 
of uniform and shellular rotation. As commented above, for such degenerate pairs, 
only the second-order degenerate coefficient is nonnull. It can be shown that the 
dominant terms correspond to $\wdabD$ and $\wdabT$ (Eqs.~\ref{ap.eq:defwdabD} and 
\ref{ap.eq:solwdabT}). The first one takes into account the distortion of the 
structure of the star through the structure quantities $d_1$ and its derivative 
$d_2$ (Eq.C20). The second one, $\wdabT$ corresponds to the toroidal component of 
eigenfunctions. For the range of frequencies treated here, the global effect remains 
small for high-frequency $p$ modes with respect to what is found in the \gp-modes 
region, where $\wdabD$ and $\wdabT$ are of the same order. This explains that 
differences between differential and uniform rotation are mainly found for 
\gp\ modes. Such variations can represent a few percent of the value of 
frequencies not corrected by near degeneracy. 

For $m=0$ modes, compensations of dominant terms turn out to yield a smaller effect 
of near degeneracy  for $m=0$ modes seen in Fig.~\ref{fig:cij2-w} (left panel), 
compared to those with $m\neq0$ (right panel).

For modes in the high frequency $p$-mode region, the near degeneracy effects shown 
in Fig.~\ref{fig:cij2-w} can be studied in the asymptotic approximation by 
analyzing individual terms of Eq.~\ref{ap.eq:defwdabD} in Appendix~\ref{sap:detail-wdabD}.  
This analysis reveals that for large frequency $p$ modes the small separation 
corresponding to $\nu_{n-1,\ell}-\nu_{n,\ell+2}$ is small enough with respect to 
the rotation rate of the star to degenerate all ($\ell,\ell+2)$ pairs. In this 
context (asymptotic regime), near degeneracy contributes with a slowly-increasing 
term with radial order, dominated by $\wdabD$ (see details in 
Sect.~\ref{ssap:w2abD-high_p}), which is given to a good approximation by
\eqn{\wdabD\sim \, \frac{\rotbar^2}{GM/R^3}~ \frac{24}{7\sqrt{5}}\,~\hat \omega_0\,<{\cal S}_2>\,,
\label{eq:w2abD-high_p_l-02}}
for ($\la,\lb$)=(0,2) degenerate pairs and
\eqn{\wdabD\sim \, \frac{\rotbar^2}{GM/R^3}~ 
\frac{12}{5}\dst\sqrt{\frac{2}{7}}\,~ \hat \omega_0\,<{\cal S}_2>\,,
\label{eq:w2abD-high_p_l-13}}
for ($\la,\lb$)=(1,3) degenerate pairs. For high-frequency $p$ modes, $C_{ab}^{(2)}$ 
increases as $\bar \omega_0$ hence the ratio $C_{ab}^{(2)}/\bar \omega_0$ remains
nearly constant.

Fig.~\ref{fig:cij2-w} also reveals that the behavior of the  $C_{ab}^{(2)}$ 
coefficients differ when differential or uniform rotation is considered. These 
differences are found significant for $m\neq0$ degenerate modes and
are mainly caused by differences at zeroth order ($\omega_0$ in 
Eqs.~\ref{eq:w2abD-high_p_l-02} \& \ref{eq:w2abD-high_p_l-13}).

For modes with $\Delta\ell=0$ (for which $C_{ab}^{(1)}\neq0$), further information 
can be obtained through the second-to-first order coefficient ratio
\eqn{q_{(2,1)}=\left|\frac{C_{ab}^{(2)}}{C_{ab}^{(1)}}\right|\,.\label{eq:defq21}}
In the present case, such degenerate pairs (which represent only a few marginal cases)
are found for $\ell=2$ and $\ell=3$, for which $C_{ab}^{(2)}\sim C_{ab}^{(1)}$. 
When considering differential rotation, mixed modes ($n=-6..2$) present second-order 
degenerate coefficients slightly higher (from 1\% to 20\%, in absolute value) than 
first order ones. Although the correction for near degeneracy is thus less important 
in the case of shellular rotation, it cannot be neglected.

\section{Discusion and Conclusions\label{sec:conclu}}

In the present work,  adiabatic oscillation frequencies of stellar models with 
intermediate mass have been computed,which include effects of rotation. Following 
the perturbation formalism of DG92 and \soufi, we have built a numerical 
code which provides adiabatic eigenfrequencies  corrected up to second order in 
the rotation rate $\Omega$ allowing a radial dependence of $\Omega=\Omega(r)$. 
Two particular cases for the transport of the angular momentum in stellar interiors 
are investigated. On the one hand, an instantaneous transport of the angular momentum 
thus leading to a uniform rotation profile. On the other hand, a 
\emph{shellular rotation} profile $\Omega(r)$ is assumed and derived from the 
assumption of local conservation of  angular momentum. Our study has been focused 
on  $1.80\msol$ equilibrium models of typical \dss\ with a surface rotation of 
$100\,\kms$.

{\bf As mentioned in introdution, seismology of A-F rapidly rotating 
stars faces many difficulties, among which, the most severe probably is mode 
identification 
\citep[see reviews][ for \ds\ and \gds\, respectively]{Goupil05,Dupret05roma}
We stress that the present study does not intend to solve the problem of mode 
identification which starts now to be handled with confidence using 
various techniques \citep[see][ and references therein]{Handler05rev}. 
In particular, diagnostics based
on multicolor photometry \citep{Garrido90} are nowadays employed for mode 
identification. In the framework of the forthcoming space mission Corot,
the colour photometric data from the exo-planet camera will provide
very useful information for this task. In any case, whether mode identification 
is successfully performed or not, calculation of models and associated 
oscillation frequencies as close to reality as possible is necessary for a one to 
one frequency comparison between observed and computed frequencies (i.e. spectrum 
of frequency differences or frequency histograms). At that level, for establishing
possible seismic diagnostics, it is important to be able to compute frequencies 
with different physical assumptions and assess their respective effects. 
The present study has been made along this viewline.}

To estimate the numerical and theoretical uncertainties of the computed
eigenfrequencies, as wel as the validity of the adopted perturbation
technique is not an easy task. Here we have performed a one-to-one eigenfrequency 
comparison between eigenfrequencies computed with the present numerical code 
and those computed with the Warsaw oscillation 
code (restricted to second-order perturbation frequency corrections).
This excersise also implies the comparison of the different perturbation techniques 
employed in each code. The input model for both codes 
is the same, $B_G$, with uniform rotation (Tab.~\ref{tab:ABmodeles}). 
The results show that differences less $0.15\,\muHz$ are most oftenly obtained and 
never larger than $0.5\,\muHz$. Both oscillation codes are different in their numerical 
schema, treatments of boundary conditions and other technical computation 
characteristics. The adopted perturbation technique is also different as, in one case, 
the first-order frequency correction is explicitly calculated (present code) whereas 
in the other case, the first-order effect of rotation is implicitly included in the 
eigensystem. The results of the comparison hence give some estimate of the precision 
of the frequencies computed here and give us some confidence in their calculation. 
   
Note also that triple-mode interaction occurs between three modes such as 
$\ell_a=\ell_b=\ell_c-2$ or $ \ell_a=\ell-b-2=\ell_c-2$ in the adopted perturbation 
formalism here, although only occasionally;  these triple-mode interactions correspond 
to a double-mode interaction in the \soufi's formalism since in this last one, 
the case $l_a=l_b$ near degeneracy is implicitly included.

Important effects -- up to $3\,\muHz$ -- of shellular rotation compared with a uniform 
rotation are observed in the low frequency region where \gp\ modes are encountered. 
This is expected as the rotation profile rapidly varies in the inner layers at 
the edge of the convective core where \gp\ modes have large amplitudes.

In the case of higher frequency $p$ modes, frequency differences up to $3\,\muHz$ are 
predicted between a uniform and a shellular rotation. This is due to the structure 
deformation caused by the centrifugal force, which mainly affects the zeroth-order 
oscillation frequencies. 
 

It is found that, the impact of reducing artificially  the core rotation rates 
of models by 50\%, principally affects \gp\ modes for which the effects shellular 
rotation are drastically reduced to $0.6$--$1\,\muHz$. In addition, the frequency 
changes due to a shellular rotation compared to a uniform rotation are found much 
larger for these intermediate rotators than third order ($\Omega^3$) effects (around 
$0.04\,\muHz$ for the region of $g$ and \gp\ modes, and around $0.02\,\muHz$ for $p$ modes, 
for a $100\,\kms$ star model). This represents around 3.3\% and 2\% of effects found 
for the \emph{core-reduced rotation} models commented here-above.

{\bf Note evenso that such effects can even be larger for different kind of stars 
(Su\'arez et al., in preparation). Observational evidences for this
have already been obtained for the early-B type stars \object{HD\,129929}
and \object{HD\,29248} \citep[][ respectively]{Aerts03science, Alosha04nueri}.}

For the models and rotation rates investigated here, quite a number of modes are 
near degenerate and a double-mode interaction had to be computed for these modes; on 
the other hand, we find that only a few modes require a triple-mode interaction 
$\ell=0, 2, 4 $ or $\ell=1, 3, 5$: one set at high frequency $n= 9,10,11$ and
two or three sets at low frequency (in the fundamental mode vicinity). For these triple 
near-degenerate modes, we encounter all possible cases of the two closest modes among 
the three being either $\ell=0,\ell=2$ or $\ell=0, \ell=4$ or $\ell=2,4$ ; this is the 
same for $\ell=1,3,5$ modes. The effect of taking into account triple interaction instead 
of double interaction can be quite different depending on the configuration of the 3 modes
(closeness of the frequencies). We find that the effect can be occasionally amount up to 
a few $\muHz$.
  
In view of our different tests as indicated above, we conclude that if one looks for seismic 
signatures of shellular rotation smaller than $0.5\,\muHz $, we have to resort to 
\soufi\ and \citet{Karami05} formalism and include triple-mode interaction 
whenever it is necessary. 
  
For higher rotation rates, perturbation techniques are no longer useful, if even still valid,
when the effects of rotation are large enough that triple-mode interaction must be 
systematically taken into account for most modes; this situation indeed would suggest that
higher than triple-mode interaction ought to be included. Then non-perturbative techniques 
such as the one developed by \citet{LigRieu04} and Rees et al.~2005 (in  prep.), for instance, 
will have to be used.

In the framework of the forthcoming space mission COROT \citep{Baglin98}, precisions of 
$0.1\,\muHz$ on mode detection, and around $0.5\,\muHz$ on splitting resolutions are expected 
to be observed. With such an accuracy, effects of shellular rotation are likely to 
detectable, provided numerical eigenfrequencies reach this level of precision. 
Finally, we stress that it is possible to remove contaminations of high-order (second 
and higher) effects of rotation from the rotational splitting in order to recover the 
true rotation profile with usual inversion techniques: this has been outlined by 
\citep{DziembowskiGoupil98,Goupil04} and will be described in more detailed in a 
forthcoming paper. {\bf At low rotation rates, this task would become even
easier, since, in this case, the effect of differential rotation on oscillation frequencies 
is dominated by structure terms in zeroth and first-order frequency corrections. }


\acknowledgements{This study would not have been possible without the financial support 
                  from the European Marie Curie action MERG-CT-2004-513610. As well, 
		  this project was also partially financed by the Spanish 
		  "Consejer\'{\i}a de Innovaci\'on, Ciencia y Empresa" from the
		  "Junta de Andaluc\'{\i}a" local government, and by the Spanish 
		  Plan Nacional del Espacio under project ESP2004-03855-C03-01. 
		  We gratefully thank W. Dziembowski for his pertinant critical 
		  remarks.}


\bibliography{/home/jcsuarez/Boulot/Latex/Util/References/ref-generale}
\bibliographystyle{aa}

\newpage


\appendix

\section{Oscillation frequencies of a pseudo-rotating model
\label{ap:oscil-pseudo}}

Considering an effective gravity, $g_{eff}$ (Eq.~\ref{eq:graveff}), 
and a rotation profile, $\Omega(r)$ (Eq.~\ref{eq:defeta0}), the following 
dimensionless quantities are defined as in \soufi:
\eqna{\you&=&\frac{\xi_r}{r}\label{ap.eq:defy01}\\
      \yod&=&\inv{\geff \,r}\big(\phi\prime
                +\frac{p\prime}{\rho}\big)\label{ap.eq:defy02}\\
      \yot&=&\frac{\phi\prime}{\geff \,r}\label{ap.eq:defy03}\\
      \yoc&=&\inv{\geff}\deriv{\phi\prime}{r}\label{ap.eq:defy04}}
Following \citet{Unno89}, we obtain a linearized eigenvalue system: 
\eqna{x\deriv{\you}{x}&=&\lambda-3\you+\frac{\Lambda}{C_r\sigocar}\yod \nonumber \\                              
      x\deriv{\yod}{x}&=&(C_r\sigma^2_0-A^*)\you+(A^*+1-\Ux)\yod-
                                  A^*\yot \nonumber \\ 
      x\deriv{\yot}{x}&=&(1-U_\chi)\yot+\yoc \label{ap.eq:sist_ordzero}  \\                                    
      x\deriv{\yoc}{x}&=&\frac{U}{1-\sigma_r}
                            \Big[A^*\you+V_g(\yod-\yot)\Big]
                            +\Lambda\yot
			    -\Ux\yoc\,, \nonumber }
with $x=r/R$ and  $R$ the stellar radius, where the following classical notations 
from \citet{Unno89} are used 
\eqna{A^*\!\!&=& \inv{\Gamma}\deriv{\ln p}{\ln r}-
               \deriv{\ln \rho}{\ln r}\label{ap.eq:defA}\\
      V&=&-\deriv{\ln p}{\ln r},~~~~~~~~~~V_g= \frac{V}{\Gamma_1}\label{ap.eq:defV}\\
      U &=&\deriv{\ln M_r}{\ln r}\label{ap.eq:defU}\\
      \Lambda&=&\ell(\ell+1)\,,}
and with $\Ux$ defined as follows:
\eqna{\Ux\! &=&  U+ \chi\,.        \label{ap.eq:defVx}}
We also define, as in \soufi, the following variables
\eqna{C &=& \Big(\frac{r}{R}\Big)^3 \frac{M}{M_r},~~~~~~~~~~~~~
      C_r = \frac{C}{1-\sigma_r} \label{ap.eq:defC}\\ 
      \sigma_0^2\!\! &=& \frac{{\omega_0}^2\,R^3}{G\,M}\label{ap.eq.defsigma0} ~~~~~ ;  
      ~~~\sigma_r\!\! = \frac{{\cal A}_c}{g} \\
      \chi &=& \frac{{\cal A}_c}{\geff}  
            \Big(U-3+\deriv{\Omega^2/\bar \Omega^2}{r}\Big)\label{ap.eq:defChi}\\
      \lambda &=& V_g(\you-\yod+\yot)\,,\label{ap.eq:deflambda}}
where ${\cal A}_c$ is the radial component of the centrifugal acceleration
defined in Eq~\ref{eq:fcentrif}; $M$, $m_r$ respectively are the stellar mass and the 
mass enclosed in the sphere of radius $r$. This system is solved with the following 
boundary conditions:
\eqna{&\!\!\!\!\yod+\you\dst\frac{\dst 3}{\dst V_g}=0,~&
        3\you+\yoc=0~~~(\ell=0)\\
       &\you-\yod\dst\frac{\dst \ell}{\dst C_r\sigma^2_0}=0,~&
        \yoc-\ell\yot=0~~~(\ell\neq0)	
      \label{ap.eq:condlim_center}}
at the center of the star and,
\eqna{&&\you = 1 \\
      &&\yoc+(\ell+1)\yot= 0  \\
      &&\you\Big(1+\dst\frac{\Lambda}{VC\sigma_0^2}-
      \dst\frac{4+C\sigma^2_0}{V}\Big)-
          \yod\big(1-\dst\frac{\Lambda}{VC\sigma_0^2}\big)+\nonumber\\
      &&  \yot\Big(1+\dst\frac{\ell+1}{V}\Big)=0
       \label{ap.eq:condlim_surf}} 
at the stellar surface. The resulting eigenvalues correspond to the 
oscillation frequencies of a pseudo-rotating model (with their associated
eigenfunctions). 

\subsection{First-order perturbed eigenfuctions \label{sap:1stordEigenf}}

Considering dimensionless variables equivalent to 
Eq.~\ref{ap.eq:defy01}--\ref{ap.eq:defy04} with first-order perturbed 
quantities ($\xi_{1,r}$, $\phi_1^\prime$, $p_1^\prime$), and the zeroth-order 
solutions obtained from Eq.~\ref{ap.eq:sist_ordzero}, first-order perturbed 
eigenfuctions can be calculated from the following system:
\eqna{x \deriv{\yu}{x} & = & \lambda_1-3\yu + \frac{\Lambda}{C_r\sigocar}\yd \\
			    &+&(\you+\zo)(1+\etao)-(\etao+ \sigma_1) \Lambda \zo 
			    	\nonumber \\
      x \deriv{\yd}{x} & = & (C_r\sigocar-A^*)\yu+(A^*+1-\Ux)\yd-A^*\yt\nonumber\\
                        & + & (\sigma_1+\eta_0)\you -(1+\eta_0) z_0\nonumber\\
      x \deriv{\yt}{x} & = & (1-\Ux)\yt+\yc \label{ap.eq:sist_ord1} \\
      x \deriv{\yc}{x} & = &  \frac{U}{1-\sigma_r} \Big[ A^* \yu + V_g \yd - V_g \yt \Big]
			    +\Lambda \yt -\Ux\yc \nonumber}
where $\lambda_1= V_g(y_1-y_2+y_3)$ and 
\eqn{\zo = \frac{\yod}{C\sigocar}\,.\label{ap.eq:defzo}}
The horizontal component of $\xiu$ can be written as follows:
\eqn{z_{1}=\frac{y_2}{C\sigma_0^2} + \frac{1+\etao}{\Lambda}y_{01}
              +\left(\frac{1+\etao}{\Lambda}-{ \sigma_1}\right)z_{0},
	      \label{ap.eq:defz1}}
where ${\sigma_1}=C_L-J_0$ represents the first-order correction
of the corresponding eigenfrequency. 
	      
\section{ $X_i$ and $Y_i$  expressions in Saio's notation
         \label{ap:saioform}}

The terms $X_i$ and $Y_i$ are constructed from $\wdI, \wdT, \wdP$ 
and, $\wdD$ (see Sect.~\ref{sec:OscFreq}). It is convenient to 
 split $X_1$ and $Y_1$ as follows:
\eqna{  X_1  &=&  X_1^I + X_1^T \nonumber \\
        Y_1  &=&  Y_1^I + Y_1^T + Y_1^P + Y_1^0 \nonumber}
These two terms include both, the toroidal component of the first 
order perturbation of the displacement eigenfunction and an effect of inertia. 
The different components of $X_1$ are given by the following
analytic expressions:
\eqna{X_1^I & = & \frac{1}{(8\Lambda-6)I_0}\int_{0}^{R}
b_2\Big[2(\Lambda-1)\you^2-\Lambda\you z_0\Big]\,\rho_0\,r^4\,\dr\nonumber\\}
\eqna{X_1^T & = & \frac{2}{I_0}\int_{0}^{R}(1+\eta_0)^2\Big[
(\you-\ell z_0)^2\frac{(\ell+1)(\ell+2)}{4(\ell+1)^2-1}\nonumber\\
&+&(\you+(\ell+1)z_0)^2\frac{\ell(\ell-1)}{4\ell^2-1}\Big]\,\rho_0\,r^4\,\dr.}
Respectively, the components of $Y_1$ are expressed as:
\eqna{Y_1^I & = &C_l-\udemi \nonumber\\
&-&\frac{1}{2I_0}\dst\int_{0}^{R} ~ \eta_2
\big[\you^2+(\Lambda-2)z_0^2-4\you z_0\big]\,\rho_0\,r^4\dr
\label{ap.eq:y1t}}
\eqna{Y_1^T & = & -\frac{2}{I_0} \int_{0}^{R}(1+\eta_0)^2\left[
(\you-\ell z_0)^2\frac{\ell+2}{\ell+1}\inv{4(\ell+1)^2-1} \nonumber \right.\\
& + & (\left.\you+(\ell+1)z_0)^2\frac{\ell-1}{\ell}\inv{4\ell^2-1}
\right]\,\rho_0\,r^4\,\dr}
\eqna{Y_1^P & = & \frac{2}{I_0}\int_{0}^{R}(1+\etao)\Big[z_0\yu +
\zu\you - \you \yu \nonumber\\
 &-& (\Lambda-1)\zu z_0\Big]\,\rho_0\,r^4\,\dr}
\eqna{Y_1^0 & = & \udemi (\sigma_1 - 1)^2.}
Furthermore, $X_2$ et $Y_2$ contain the 
effect of deformation of the star due to the non-spherical component of the 
centrifugal acceleration. They are given by:
\eqn{X_2 =  \frac{\Lambda}{4\Lambda-3}\,{\cal I}_c  ~~~~~~~~~~~~~~~
Y_2  =  \frac{-3}{4\Lambda-3}{\cal I}_c,}
with
\eqna{{\cal I}_c & = &
\inv{2I_0}\dst\int_{0}^{R}\Big\{C\sigocar\big[(2\you^2+3z_0^2)
\dst(r\deriv{u_2}{r} +(4-U)u_2)\nonumber \\
& + & 2z^2_0\Lambda u_2+z_0\you\dst(\Psi_1+6u_2)\big]\nonumber \\
& + & \you\big(\lambda-\you(A^*+V_g)\big)Uu_2-\you\yot(\Psi_1+2\Lambda u_2)\nonumber \\
& - & \you\yoc\big(3r\deriv{u_2}{r}+u_2(10-3U)\big) \label{ap.eq:defIc_1}\\
& - & 2z_0\yoc\Lambda u_2-\big[\you^2(C\sigocar+3-U)+z_0^2C\sigocar(\Lambda-3)\nonumber \\
& + & 3z_0\you-2\you\yoc-2z_0\yot(\Lambda-3)\big]b_2\nonumber \\ 
& - & \you^2r\deriv{b_2}{r}\Big\}\,\rho_0\,r^4\,\dr ,\nonumber}
with $\ud$, $\bd$, $\eta_2$ defined as in \dg:
\eqna{\ud&=&\frac{\phi_{22}}{r^2\rotbar^2}+\utier(1+\etad)\nonumber\\
     \bd &=& \inv{3}r\deriv{\etad}{r}\label{ap.eq:defu2}   \\
      \eta_2 &=& (1+\eta_0)^2 -1\,.  \nonumber } 
%
\subsection{High radial order $p$ modes\label{sap:hraopmodes}}
 
For high radial order $p$ modes, the horizontal component of the fluid 
displacement is much smaller than the vertical one and $ |y_{01}| >> |z_0|$. 
The modes are concentrated toward the surface and 
$y_{03},y_{04} \sim 0$, $U\sim 0$, $C\sim 1$ , $\lambda\sim V_g y_{01}$. 
Then Eq.~\ref{ap.eq:defIc_1} becomes:
\eqna{{\cal I}_c & \sim &
\inv{2I_0}\dst\int_{0}^{R} ~ y^2_{01} ~
 \Big\{C\sigocar \big[(2 \dst(r\deriv{u_2}{r} + 4 u_2)-  b_2\big] \nonumber \\
& - &  (C\sigocar +3) b_2 -r\deriv{b_2}{r} \Big\}\, \rho_0\,r^4\,\dr\,.}
For large $\sigma_0$, one further has
\eqna{{\cal I}_c & \sim &
\sigocar\inv{I_0}\dst\int_{0}^{R} ~ y^2_{01} ~ 
  ~\Big\{ \dst(r\deriv{u_2}{r} + 4 u_2-b_2 \Big\}\,\rho_0\,r^4\,\dr\,. }
As 
$$I_0 \sim  \dst\int_{0}^{R} ~\, y^2_{01} ~ \rho_0\,r^4\,\dr, $$
then
\eqn{{\cal I}_c \sim \sigma^2_0 < {\cal S}_2 >,\label{ap.eq:defIavecS2}}
with
\eqna{< {\cal S}_2> & = & < \Bigl(r\dst\deriv{u_2}{r} + 4 u_2-b_2\Bigr) >,\label{ap.eq:defS}}
where $<>$ is a  weighted average defined in Eq.~\ref{eq:defaverage}.  
Hence $ {\cal I}_c$ (thereby $X_2, Y_2$) increases with $\sigma_0^2$ i.e. with 
increasing frequency.

\subsection{Radial modes \label{sap:rmodes}}

For radial modes, $\ell=0$, one has  $X_2=0$ Eq.~\ref{eq:w2tot_saio}  simplifies as:
\eqn{\omega_{2} = \frac{\rotbar^2}{I_0\,\omega_0}
                    \dst\int_{0}^{R}  \,  \you^2\,\dst\Big(b_2+\frac{4}{3} (1+\eta_0) \Big) \,
		    \rho_0\,r^4\,\dr
      \label{ap.eq:w2tot_saio_lzero}}

\section{First- and second-order corrections for near-degenerate frequencies  
         \label{ap:terms2ord}}

The different terms which contribute to the first- and second-order frequency corrections in 
presence of degeneracy are collected in the $\muj$ and  $\mujk$  given by 
Eq.~\ref{eq:Defmujk}. These contributions are :
\eqna{\wuab&=&-\inv{I_a}\xioabra\,\ka\,\xiobket\label{ap.eq:wuab-inic}\\ 
\wdabPT &=&-\dst\inv{I_a}\Big[\xioabra\,\ka\,\xiubpket+\xioabra\,\ka\,\xiubtket\Big]\nonumber\\
&=& \wdabP+\wdabT \label{ap.eq:wdabT-torpol}\\
\dst\wdabD &=&\inv{2 \bar \omega_0 I_a}\,
\dst\xioabra \frac{1}{\rho_0}\left( \eld-\rhod \, \bar \omega_0^2 \right)\xiobket
\label{ap.eq:w2abdef}\\
\wdabI&=& \inv{2 \bar \omega_0 I_a}\Big[\dst\xioabra\,m^2{\bs\Omega^2}
-\,2\,m{\bs\Omega}\xiobket\Big],}
where $I_a$ (Eq.~\ref{eq:defI0}) is the normalization term for the mode $a$,
\eqn{I_a=\xioabra\xioa\rangle,}
where the scalar product definition is defined as in  Eq.~(12) of \dg.
$\xiukP$ and $\xiukT$ represent respectively the poloidal and toroidal 
components of first-order eigenfunction correction (see Eq.~26 in \soufi\ for 
explicit expressions). The operator $\ka$ is defined as:
\eqn{\ka = m{\bs\Omega}-\i{\bs\Omega}\ez\times\,.\label{ap.eq:defKappa}}
Definitions of the operators $\eld ,\rho_2$ can be found in 
\dg\ and \soufi. 

\subsection{Detailed expression for $\wuab$ \label{sap:firstterms}}

The use of the definitions of $\eld$ and $\ka$ on
Eqs.~\ref{ap.eq:wuab-inic}--\ref{ap.eq:w2abdef} yields the
following expanded expressions:
\eqna{\wuab&=& \frac{m\rotbar}{I_a}\kro{\la}{\lb}\dst\intr r^4 
\rhoo\dr\,  (1+\eta_0) \, \Big\{\nonumber\\
&& -\big(\youa\youb+\delb\zoa\zob\big)\nonumber\\
&& +\big(\youa\zob+\zoa \youb+ \zoa \zob \big)
\Big\},\label{ap.eq:solwuab}}
where $\kro{\la}{\lb}$ is the Kronecker symbol. 
%

\subsection{Detailed expression for  $\wdabPT$ and $\wdabI$ }

In the case of $\wdabPT$, its  components are expressed as: 
\eqna{\wdabP&=& - 2 \frac{m^2\rotbar^2}{ I_a \bar \omega_0 }\kro{\ell_a}{\ell_b} 
\dst\intr r^4\rhoo\dr \,(1+\eta_0)^2\,  \Big \{ \\
&& y_{01,a} \yub+\dela z_{01,a} \zub \nonumber \\
&& - ( y_{01,a}\zub+ z_{01,a} \yub+ z_{01,a} \zub) 
\Big\}\,,\label{ap.eq:solparpol}}
and
\eqna{&&\wdabT=\frac{2\rotbar^2}{I_a \bar \omega_0 }
\intr  r^4\rhoo\dr  (1+\eta_0)^2 \,\Big \{\nonumber\\
&&\delta_{\ell_a,\ell_b}\big( \tau_{b+1}\tau_{b+1}\Lambda_{\ell_b+1}
+ \hat \tau_{b-1}\hat \tau_{b-1}\Lambda_{\ell_b-1} \big) \\
&&+ \kro{\la}{\lbpd} \,\big(\tau_{b+1}\hat \tau_{b+1} \,\Lambda_{b+1}\big)
+ \kro{\la}{\lbmd} \big(\tau_{b-1}\hat \tau_{b-1} \,\Lambda_{b-1} \big) \Big\},\nonumber 
\label{ap.eq:solwdabT}}
where $\tau_{b+1}$ and $\hat \tau_{b-1}$ are defined as follows:
\eqna{\tau_{b+1} &=&
-\frac{\i}{\left(\lb+1\right)} \left(\youb-\lb\zob\right)\gambp\\
\hat \tau_{b-1}
&=&\frac{\i}{\lb} \left(\youb-(\lb+1)\zob\right)\gamb\,.\label{eq:defttau}}
Finally, the \emph{inertial} term of the correction is given by 
\eqna{\wdabI&=& -m^2 \,\frac{\rotbar^2}{2  I_a \bar \omega_0}\,\kro{\la}{\lb} 
\, \intr \,r^4\rhoo\dr\,(1+\eta_0)^2 \,\Big \{\nonumber\\
&&  \big(\youa\youb+\delbp\zoa\zob\big)\nonumber\\
&& -2\,\big(\youa\zob+\youb\zoa+\zoa\zob\big) \Big\}\,.
    \label{ap.eq:solwdabI}}
%

\subsection{Detailed expression for $\wdabD$ \label{sap:detail-wdabD}  }

For the correction $\wdabD$, the angular dependence can be factorized as follows:
\eqn{\wdabD =\frac{\rotbar^2}{\bar \omega_0}~ \qdab\,~ \zab,\label{ap.eq:defwdabD}}
where $\qdab$ and $\zab$ represent the angular and radial components
respectively. Selection rules then exist and are imposed by:
\eqn{\qdab =
\intang\,\yac\yb\,P_{2}(\cost)\dang,\label{ap.eq:defQdab}}
where $\dang= \sin \theta d\theta d\varphi$ is the elementary solid angle 
and $ \yb$ is the spherical harmonics $Y_{\ell_b,m_b}$.
Using the second-order Legendre polynomial:
\eqn{P_2(\cost)=\frac{3}{2}\,\cosdt-\inv{2},\label{ap.eq:defP2}}
we  obtain
\eqna{\qdab&=&\frac{3}{2}\Big[
\big(\gamma^2_{a+1}+\gamma^2_a-\inv{3}\big)\kro{a}{b}\nonumber\\
&&+\gama\gambp\kro{a}{b+2}+\gamb\gamap\kro{a}{b-2}\Big],}
with
\eqna{
&\gamb&=\sqrt{\Fb} ~~~~~~~~{\rm and} ~~~~\Fl=\frac{\ell^2-m^2}{4\,\ell^2-1}.}

On the other hand, the radial component $\zab$ can be written as 
\soufi:
\eqn{\zab=\inv{2I_a}\intr
r^2\rhoo\dr\Big\{\du\Eu+\dd\Ed+r^2\bd\Et+r^2\bt\Eq\Big\},\label{ap.eq:zabE1234}}
where 
\eqn{\du~ = r^2\ud~~~~ \dd=r\deriv{\du}{r} ~~~~~~~  \bt = \frac{1}{3} r^2 \deriv{^2\eta_2}{r^2}\,,
\label{ap.eq:Defduddbdbtbtp}}
with $b_2$ defined in B9. The non-degenerate expression is a particular case when $a=b$ and it can be
found in \dg.

The explicit expressions of $\Eu$, $\Ed$, $\Et$ and $\Eq$ are given below.
For the sake of shortness, the following nomenclature is used in  \dg:
\eqna{y&\equiv&\you ~~~~~~v\equiv\yot ~~~~~~~~W\equiv\yoc\,,
      \label{ap.eq:WojFilequiv}}
and the following \emph{short expressions} have been defined:
\eqna{q_j&=&y_j\left(C\sigma^2_{0,j}+4-U\right)-\Lambda_jz_{0,j}-W_j
     \label{ap.eq:defq}\\
     \lambda_j&=&\vg\left(y_{j}-y_{02,j}+v_j\right)\label{ap.eq:deflambda_ij}\\
      s_j&=&\frac{p\,\Gamma_1\lambda_j}{r\rho\,\geff}.\label{ap.eq:defs}}
for $j=a,b$. We then have:

\eqn{E_k=E_{k,ab}+E_{k,ba}\, ~~~~~~~~~~~~~~~~~~k=1,4 }
with 
\eqna{
\Eu_{ab}&=&-\frac{3}{2}\qa\fb+ \frac{1}{2}\fa\fb{\cal F}_{1}-\qa \lambb -\inv{2}\fb s_a \psi \nonumber\\
&+&(C\sigma^2_{0,a}+4)\fa \lambb-\lambb\wa\nonumber\\
&+&\left(U-\frac{3}{2}\right)\wa\fb-\dela\va\fb\nonumber\\
&-&\dela\zoa\wb+3\qa\zob+ \frac{1}{2} \zoa\zob{\cal F}_{3}\nonumber\\
&-&\dela\zoa\fb\left(\frac{3}{2}- C\sigma^2_{0,a}+ U-4\right)-\dela\zoa\lambb\nonumber\\
&-&\zoa\fb{\cal F}_{2b}+\zoa\yodb(\delabp-3)(U-4)\nonumber\\
&+&\inv{2}(4-U) \Bigl(\fa \qb- s_a \Lambda_b z_{0b}\Bigr),\label{ap.eq:result-E1num}
}
\eqna{\Ed_{ab}&=& \frac{1}{2}(C\sigma^2_{0,a}-U)\fa\fb+\inv{2}\fa \qb+(U-2) \fa s_b\nonumber\\
&-&\fb\wa+\frac{1}{2}(\delabp-3)\,C\sigma^2_{0,a}\zoa\zob+ \Lambda_b z_{0b} y_a\nonumber\\
&-&(\delabp-3)(\zob\yoda)-\inv{2} s_a\Lambda_b z_{0b},\label{ap.eq:result-E2num}}
\eqna{\Et_{ab} &=&-\frac{1}{2} C\sigma^2_{0,a} \left(\fa\fb+(\delabp-3)\zoa\zob\right)\nonumber\\
&-&\inv{2}\fa s_b (U+6)-3 \fb\zoa - \inv{2}(\dela-\delb)\fb\zoa\nonumber\\
&-&(3-\delabp)\zoa\vb+\fa\wb\nonumber\\
&+&\inv{2}s_a \lambda_b\left(\derivplnln{\gamu}{r}\right)_{\po},
\label{eq:result-E3num}}
\eqna{\Eq&=& -\inv{2}\ysab\label{ap.eq:defE44}\label{eq:result-E4num},}
with
\eqna{{\cal F}_{1}&=&6+U(U-2(\aet+\vg)-3)+(3-U) C\sigma^2_{0,a} \nonumber\\
{\cal F}_{2a}&=&(\delabp-3)(U-4)+\delabp (C\sigma^2_{0,a}+4-U)\nonumber\\
{\cal F}_{3}&=&(\delabp-3)C\sigma^2_{0,a}(6-U)+6\delabp\\
\psi&=&(1-U)(4-U)+6+U(U-3) \nonumber \\
\delabp&=&\inv{2}\left(\dela+\delb\right)\label{ap.eq:delabp}\nonumber\\
\delbam&=&\inv{2}\left(\delb-\dela\right)=-\delabm ,
\label{ap.eq:defE1}}
where
\eqn{\Lambda_{j}=\ell_j(\ell_j+1),\label{ap.eq:defdela}}
with $j$ representing the modes $a$ and $b$.

\subsubsection{High radial order $p$ modes \label{ssap:w2abD-high_p}}

Following the same approximations for modes concentrated towards the surface as in Sec.~\ref{sap:hraopmodes},
and using the following approximations (from the definitions 
Eqs.~\ref{ap.eq:defq}--\ref{ap.eq:defs}) 
\eqna{q_j&\sim &y_j\left(C\sigma^2_{0,j}+4\right) , \, \lambda_j\sim \vg y_{j}, \,  s_j\sim  y_{j}  ~~~j=a,b}
Eq.~\ref{ap.eq:zabE1234} becomes for large $\sigma_0$: 
\eqna{\zab &\sim& \inv{I_a}\intr \,C\sigma_0^2  
\Big[  4 u_2 +  r\frac{d u_2}{dr} -  b_2 \Big] y_a y_b~   \rhoo r^4  \dr \, ,
\label{ap.eq:zab_asympt_1} }
where we have used definitions given in Eq.~\ref{ap.eq:defu2} and 
Eq.~\ref{ap.eq:Defduddbdbtbtp}. This expression is the same as the one in B11 when $a=b$.

Using the definition of weighted average given 
in Eq.~\ref{eq:defaverage} with $y_a y_b$ instead of $y_{01}^2$, 
Eq.~\ref{ap.eq:zab_asympt_1} can be rewritten as :
\eqn{\zab\sim \,\sigma_0^2\, <{\cal S}_2>_{ab}\,.}
Finally, $\wdabD$ (Eq.C14) for high radial order $p$ modes can be 
written as
\eqn{\wdabD\sim \,\frac{\rotbar^2}{GM/R^3}~ \qdab\,~\omega_0\,<{\cal S}_2>_{ab} .\label{ap.eq:wdabD-asymp}}

\end{document}